\documentclass[11pt,ruled,lined,linesnumbered,commentsnumbered,longend]{article}
\usepackage[latin9]{inputenc}
\usepackage{mathtools}
\usepackage{algorithm2e}
\usepackage{amsmath}
\usepackage{overcite}

\makeatletter
\usepackage{graphicx}
\usepackage{mathbbol}
\usepackage{xcolor}
\usepackage{xfrac}

\newcommand*\xbar[1]{%
  \hbox{%
    \vbox{%
      \hrule height 0.5pt % The actual bar
      \kern0.5ex%         % Distance between bar and symbol
      \hbox{%
        \kern-0.1em%      % Shortening on the left side
        \ensuremath{#1}%
        \kern-0.1em%      % Shortening on the right side
      }%
    }%
  }%
}

\makeatother

\title{An adaptive Bayesian approach to gradient-free global optimization}

\author{Jianneng Yu$^{1,2}$ and Alexandre V. Morozov$^{1,2}$ \footnote{Corresponding author: \texttt{morozov@physics.rutgers.edu}} \\
        \small{$^1$ Department of Physics and Astronomy, Rutgers University, Piscataway, NJ 08854, USA} \\
        \small{$^2$ Center for Quantitative Biology, Rutgers University, Piscataway, NJ 08854, USA}
        }

\date{}

\begin{document}

\maketitle

\begin{abstract}
Many problems in science and technology require finding global minima or maxima of various objective functions.
The functions are typically high-dimensional; each function evaluation may entail a significant computational cost.
The importance of global optimization has inspired development of numerous heuristic algorithms based on analogies with
physical, chemical or biological systems. Here we present a novel algorithm, SmartRunner, which
employs a Bayesian probabilistic model informed by the history of accepted and rejected moves to make a decision about the next random trial.
Thus, SmartRunner intelligently adapts its search strategy to a given objective function and moveset, with the goal of maximizing fitness gain (or energy loss) per 
function evaluation. Our approach can be viewed as adding a simple adaptive penalty to the original objective function, with SmartRunner performing hill ascent or descent
on the modified landscape. This penalty can be added to many other global optimization algorithms.
We explored SmartRunner's performance on a standard set of test functions, finding that it compares favorably against several widely-used alternatives:
simulated annealing, stochastic hill climbing, evolutionary algorithm, and taboo search. Interestingly, adding the adaptive penalty to the first three of these algorithms considerably enhances their performance.
%In the cases of simulated annealing and stochastic hill climbing, the globally best strategy turns out to reduce to hill ascent or descent -- the original SmartRunner approach.
We have also employed SmartRunner to study the Sherrington-Kirkpatrick (SK) spin glass model and Kauffman's NK fitness model -- two NP-hard problems characterized by
numerous local optima. In systems with quenched disorder, SmartRunner performs well compared to the other global optimizers.
Moreover, in finite SK systems it finds close-to-optimal ground-state energies averaged over disorder.
%We have also used SmartRunner to compute known SK ground-state energies averaged over disorder as a function of the number of spins.
%SmartRunner finds optimal or nearly optimal solutions in these systems.
\end{abstract}

\section*{Introduction}

% Discuss importance of the problem
Many models in fields of enquiry as diverse as natural and social sciences, engineering, machine learning, and quantitative medicine are described by complex non-linear functions of
many variables. Often, the task is to find globally optimal solutions of these models, which is equivalent to finding global minima or maxima of the corresponding model functions.
The global optimization problem arises in engineering design, economic and financial forecasting, biological data analysis, potential energy models in physics and chemistry, robot design and manipulations,
and numerous other settings. Notable examples include finding the minimum of protein free energy in computer simulations of protein folding~\cite{Onuchic:2004,Dill:2008}, finding high-fitness solutions in evolving
populations subject to mutation, selection, recombination, and genetic drift~\cite{Crow1970,Kimura1983,Gillespie2004} (biological fitness quantifies the degree of reproductive success of an organism in an evolving population), and minimizing the error function in deep-learning neural network models~\cite{Goodfellow2016,Mehta2019}. 

Mathematically, the global optimization problem is defined as finding the maximum (or the minimum) of a real-valued function $\mathcal{F} (X)$, where $X$ denotes a collection of discrete or continuous variables
that describe the state of the system. The states of the system may be subject to non-linear constraints.
Here we focus on maximizing $\mathcal{F} (X)$, which we will refer to as the fitness function; with $\mathcal{F} (X) = -E (X)$, this is equivalent to minimizing an energy or 
error function $E (X)$. In the energy function case, $E (X)$ may signify the energy of a microstate or a free energy of a coarse-grained/mesoscopic state.
The number of variables in $X$ may be large in real-world applications and $\mathcal{F} (X)$ may be costly to evaluate, making it highly desirable
to develop efficient global optimization algorithms
which require as few fitness function evaluations as possible to reach high-quality solutions. The set of fitness values assigned to all states of the system forms a fitness landscape -- a high-dimensional
surface which global optimization algorithms must traverse on their way to the mountain peaks that correspond to high-scoring solutions.

If the fitness function is concave everywhere, the fitness landscape consists of a single peak and the global maximum is easy to find. However, in most problems of interest fitness landscapes
contain multiple local maxima and saddle points which can trap the optimizer. There is no guarantee of finding the global maximum in this case unless all system states can be examined, which is usually
not feasible because their number is exponentially large. A well-known worst-case scenario is a ``golf-course'' landscape which is flat everywhere apart from a few states that form a basin of attraction
for an isolated deep hole, or a tall peak. In protein folding, this scenario is known as Levinthal's paradox~\cite{Zwanzig:1992} --
proteins cannot fold on biologically reasonable time scales if they need to sample a sizable fraction of their microscopic configurations. While Levinthal's paradox has been resolved by
introducing the concept of a protein folding funnel~\cite{Bryngelson:1995,Dill:1997,Onuchic:2004,Dill:2008},
generally there is no guarantee of finding the global maximum in a reasonable number of steps, and global optimization is demonstrably an NP-hard problem~\cite{Danilova2022}.

If the gradient of the fitness function can be computed efficiently, it should be used to guide the search because the gradient vector indicates the direction of the steepest ascent.
Here, we focus on systems with discrete or discretized states and assume that the gradient is not available.
Namely, we consider an undirected graph with $N$ nodes or vertices, where $N$ is the total number of system states which may be astronomically large or even unknown.
Each node $i = 1 \dots N$ is assigned a state $X_i$ and a corresponding fitness value $\mathcal{F} (X_i)$.
%With $\mathcal{F}_i = - E_i$, where $E_i$ is either the energy or the free energy of state $i$ (energy for true microstates, free energy for coarse-grained/mesoscopic states),
%our definition extends to a vast class of molecular and solid state systems in which energy minimization is desirable, for example to find folded protein states or low-energy states of atomic
%clusters.
This definition describes a vast number of systems that are either naturally discrete (e.g., spin glasses~\cite{Sherrington:1975}) or discretized by superimposing a lattice on a continuous landscape.
Besides the fitness function, a global optimization algorithm requires a move set -- a deterministic or stochastic rule for moving between states
on the fitness landscape. A move set defines state neighborhoods -- a set of states reachable from a given state in a single jump. The size of the neighborhood is typically fixed but may also
change in complex ways, e.g. with recombination moves described below.

% Discuss various algorithms

Numerous empirical approaches have been developed over the years to tackle the problem of gradient-free optimization.
Usually, these algorithms are based on an analogy with a physical, chemical or biological process in which some kind of optimization is known to occur.
For example, the celebrated simulated annealing algorithm~\cite{Kirkpatrick:1983} is a Monte Carlo technique
based on an analogy with a physical annealing process in which the material starts at a high temperature to enable constituent molecules or atoms to move around. The temperature is gradually decreased, allowing the material to relax into low-energy crystalline states. The rate of temperature decrease is a key parameter of the
simulated annealing algorithm~\cite{Cohn:1999}.
Numerous modifications of the basic simulated annealing approach have been developed over the years:
parallel tempering Monte Carlo~\cite{Hukushima:1996}, replica Monte Carlo~\cite{Swendsen:1986}, population annealing~\cite{Wang:2015b},
simulated tempering~\cite{Marinari:1992}, and many others.
Generally speaking, the idea of these algorithms is to overcome free energy barriers by simulating a broad range of temperatures.
Besides estimating various thermodynamic quantities by Monte Carlo sampling, some of these algorithms have also been applied to combinatorial optimization problems such as
the search for the ground states of Ising spin glasses~\cite{Wang:2015a}.

Genetic or evolutionary algorithms~\cite{Goldberg:1989,Vikhar2016,Slowik:2020} are based on an analogy with the evolution of a biological population: a population of candidate solutions is subjected to
multiple rounds of recombination, mutation, and selection, enabling ``the survival of the fittest''. Harmony search is a music-inspired algorithm, applying such concepts as playing a piece of music from memory, 
pitch adjustment, and composing new notes to an evolving population of harmonies~\cite{Geem:2001,Lee:2005}. Particle swarm algorithms draw their inspiration from the collective behavior of bird flocks and
schools of fish~\cite{Kennedy:1995,Eberhart:1995}. Taboo search is a deterministic strategy in which all nearest neighbors of the current state are examined and the best move is accepted~\cite{Cvijovic:1995}.
To avoid returning to previously examined states via deterministic cycles, a fixed-length ``taboo'' list is kept of the recently visited states that are temporarily excluded from the search.
Stochastic hill climbing employs a procedure in which the moves are accepted or rejected using a sigmoid (two-state) function with a fixed temperature $T$~\cite{Juels1995}. As in simulated annealing,
this strategy allows for deleterious moves whose frequency depends on the value of $T$.
Many other heuristic algorithms and variations of the above algorithms are available in the literature~\cite{Torn:1989,Berg:1993,Hesselbo:1995,Dittes:1996,Barhen:1997,Wenzel:1999,Hamacher:2006}.

Here we propose a novel global optimization algorithm which we call SmartRunner. SmartRunner is not based on an analogy with a physical, chemical or biological system.
Instead, the algorithm uses previously accumulated statistics on rejected and accepted moves to make a decision about its next move.
Thus, SmartRunner adapts its search strategy intelligently as a function of both local and global landscape statistics collected earlier in the run, with the goal of maximizing
the overall fitness gain. Generally speaking, SmartRunner can be viewed as a stochastic extension of the Taboo search policy. However, unlike the Taboo algorithm, it does not need to
evaluate fitness values of every neighbor of the current state, which may be computationally expensive. Moreover, it replaces infinite penalties assigned to the states in the ``taboo'' list by
node-dependent penalties which only become infinite when all the nearest neighbors of the node is question have already been explored. We benchmark SmartRunner on a set of challenging
global optimization problems and show that it consistently outperforms several other state-of-the-art algorithms. Moreover, we demonstrate that the SmartRunner approach amounts to hill climbing on a
dynamically redefined fitness landscape. This redefinition can be used to enhance the performance of many other global search approaches such as simulated annealing or evolutionary algorithms.

\section*{Materials and Methods}

%\subsection*{Theory}

\noindent
\textbf{Bayesian estimation of the probability to find a novel beneficial move.}

\noindent
\textit{Unweighted moves.}
Consider a fitness landscape with a move set that defines $\mathcal{N}$ nearest neighbors for each discrete system state $X_i$ $(i = 1 \dots N)$.
%The neighborhood of each node is fully defined by the move set.
%We assume that $\mathcal{N}$ is known and that it does not change with time or the node index $i$.
We divide all neighbors of the state $X_i$ into two disjoint subsets:
one set $S^{i}_{p}$ of size $U^{i}_{p} \ge 0$ contains all states with fitness $\le \mathcal{F}_i$, while the other set $S^{i}$ of size $U^{i} = \mathcal{N} - U^{i}_{p} \ge 0$
contains all states with fitness $> \mathcal{F}_i$.
%Note that by construction $U^{i}_{p} + U^{i} = \mathcal{N}$.
Moves between $X_i$ and any state in the set $S^{i}_{p}$ are deleterious or neutral, while 
moves to any state in the set $S^{i}$ are beneficial. Generally, we expect the size of $S^{i}$ to be small: $U^{i} \ll \mathcal{N} \simeq U^{i}_{p}$ because as a rule it is
more difficult to find a beneficial move than a deleterious or neutral one.

We assign the system state $X_i$ to the node $i$ on a network, with nodes representing system states and edges representing nearest-neighbor jumps.
We consider a single random walker that explores the network. %fitness landscape by jumping from node to node according to the pre-defined move set.
At each step, the walker is equally likely to initiate a jump to any of the $\mathcal{N}$ neighbors of the current node. 
Let us say that the random walker is currently at node $i$ and has made $n$ unsuccessful
attempts to make a move $i \to j \in \text{nnb} (i)$, where $\text{nnb} (i) = S^{i}_{p} \cup S^{i}$ is a set that contains all the nearest neighbors of node $i$
(for simplicity, let us assume for the moment that all deleterious and neutral moves are rejected while a beneficial move, once found, is immediately accepted).
After $n$ trials, we have data $\mathcal{D} = \{ K_p, m_p, K, m \}$, where $K_p$ is the total number of visits to the nodes in $S^{i}_{p}$
and $K = n - K_p$ is the total number of visits to the nodes in $S^{i}$. Furthermore, $m_p \le K_p$ and $m \le K$ are the number of \emph{unique} visited nodes in
$S^{i}_{p}$ and $S^{i}$, respectively. The probability of observing $\mathcal{D}$ is given by
\begin{equation} \label{eq:P:DU}
P(\mathcal{D} | U^{i}) = \binom{n}{K} \left( \frac{U^{i}}{\mathcal{N}} \right)^K \left( 1 - \frac{U^{i}}{\mathcal{N}} \right)^{n-K}.
\end{equation}
Correspondingly, the probability of $U^{i}$ given the data is
\begin{equation} \label{eq:P:UD}
P(U^{i} | \mathcal{D}) = \frac{P(\mathcal{D} | U^{i}) P(U^{i})}{\sum_{U'=0}^{\mathcal{N} - m_p} P(\mathcal{D} | U') P(U')},
\end{equation}
where $P(U)$ is the prior probability that there are $U$ nearest neighbors of node $i $ whose fitness is higher than $\mathcal{F}_i$. Choosing an uninformative prior, we obtain:
\begin{equation} \label{eq:prior}
%P(U) = \frac{1}{\mathcal{N} - m_p +1}.
P(U) = \frac{1}{\mathcal{N} + 1}.
\end{equation}
Note that $\sum_{U=0}^{\mathcal{N}} P(U) = 1$. Then Eq.~\eqref{eq:P:UD} yields
\begin{equation} \label{eq:P:UD:mod}
P(U^{i} | \mathcal{D}) = \frac{1}{Z} \left( \frac{U^{i}}{\mathcal{N}} \right)^K \left( 1 - \frac{U^{i}}{\mathcal{N}} \right)^{n-K},
\end{equation}
where $Z = \sum_{U'=0}^{\mathcal{N} - m_p} \left( \frac{U'}{\mathcal{N}} \right)^K \left( 1 - \frac{U'}{\mathcal{N}} \right)^{n-K}$.

Focusing on the $K=0$, $m=0$ limit (that is, on the case where no beneficial moves have yet been found) and assuming $U^{i} \ll \mathcal{N}$, we obtain
\begin{equation} \label{eq:P:DU:simple}
P(\mathcal{D} | U^{i}) = \left( 1 - \frac{U^{i}}{\mathcal{N}} \right)^{n} \simeq e^{-\gamma U^{i}},
\end{equation}
where $\gamma = n/\mathcal{N}$. Furthermore,
\begin{equation} \label{eq:Z}
%Z \simeq \sum_{U'=0}^{\mathcal{N} - m_p} e^{-\gamma U'} = \frac{1 - e^{-\frac{\mathcal{N} - m_p + 1}{\mathcal{N}}n}}{1 - e^{-\frac{n}{\mathcal{N}}}},
Z \simeq \sum_{U'=0}^{\mathcal{N} - m_p} e^{-\gamma U'} = \frac{1 - e^{-\gamma \widetilde{\mathcal{N}} }}{1 - e^{-\gamma}},
\end{equation}
where $\widetilde{\mathcal{N}} = \mathcal{N} - m_p + 1$.
Note that the exponential substitution becomes inaccurate for the terms in the sum in which $U'$ approaches $\mathcal{N} - m_p$; however, since $m_p \le \mathcal{N}$,
these terms are suppressed in the $n \gg 1$ limit compared to the accurately approximated terms with $U' \ll \mathcal{N} - m_p$.
We observe that $m_p$ is a stochastic variable whose expectation value can be shown to be
\begin{equation} \label{E:mp}
E[m_p] = \mathcal{N} \left[ 1 - (1 - \frac{1}{\mathcal{N}})^n \right] \simeq \mathcal{N} \left[ 1 -  e^{-\gamma} \right],
\end{equation} 
where the last approximation requires $\mathcal{N} \gg 1$.

Finally,
\begin{equation} \label{eq:P:UD2}
P(U^{i} | \mathcal{D}) = \frac{1}{Z} e^{-\gamma U^{i}} = e^{-\gamma U^{i}} \frac{1 - e^{-\gamma}}{1 - e^{-\gamma \widetilde{\mathcal{N}}} }.
\end{equation}

If $\mathcal{N} \gg 1$ and $\mathcal{N} \gg m_p$, Eq.~\eqref{eq:P:UD2} yields
\begin{equation} \label{eq:P:Uzero}
P(U^{i} = 0 | \mathcal{D}) \simeq \frac{1 - e^{-\frac{n}{\mathcal{N}}}}{1 - e^{-n}} \simeq 1 - e^{-\frac{n}{\mathcal{N}}},
\end{equation}
where the last approximation is valid for $n \gg 1$. Thus, the probability to find beneficial moves, $P(U^{i} > 0 | \mathcal{D}) \simeq e^{-{n}/{\mathcal{N}}}$, decreases exponentially
with $n$. Note that if $n=0$ (no random trials have been made), Eq.~\eqref{eq:P:UD2} yields $P(U^{i} > 0 | \mathcal{D}) = \mathcal{N}/(\mathcal{N} + 1)$, consistent with
the prior probability in Eq.~\eqref{eq:prior} which assigns equal weights to all $\mathcal{N} + 1$ values of $U^{i}$.
Thus, to begin with the system is very optimistic that a beneficial move will be found. However, if $m_p = \mathcal{N}$ (that is, all moves have been tried and none are beneficial), 
Eq.~\eqref{eq:P:UD2} yields $P(U^{i} > 0 | \mathcal{D}) = 0$, as expected. Thus, the system gradually loses its optimism about finding a beneficial move as it makes more and more unsuccessful
trials.

Finally, we compute the probability of finding a higher-fitness target in the next step:
\begin{equation} \label{eq:pf}
p_f = \sum_{U^{i}=0}^{\mathcal{N} - m_p} \frac{U^{i}}{\mathcal{N}} P(U^{i} | \mathcal{D}) = \frac{1}{\mathcal{N}} \frac{e^{-\gamma} - \widetilde{\mathcal{N}} e^{-\gamma \widetilde{\mathcal{N}}} +
e^{-\gamma (\widetilde{\mathcal{N}} + 1)} (\widetilde{\mathcal{N}} - 1)}{(1 - e^{-\gamma})(1 - e^{-\gamma \widetilde{\mathcal{N}}})}.
\end{equation}
%where $\widetilde{\mathcal{N}} = \mathcal{N} - m_p + 1$ and $\gamma = n/\mathcal{N}$.
In the beginning of the search, $n \ll \mathcal{N}$ and, correspondingly, $m_p \ll \mathcal{N}$.
If, in addition, $n \gg 1$ (which implies $\mathcal{N} \gg 1$), Eq.~\eqref{eq:pf} simplifies considerably:
\begin{equation} \label{eq:pf:limit1}
p_f \simeq \frac{1}{\mathcal{N}} \frac{1 - n/\mathcal{N}}{n/\mathcal{N}} = \frac{1}{n} \left[ 1 + \mathcal{O} (\frac{n}{\mathcal{N}}) \right].
\end{equation}
Note that in this limit $p_f$ is independent of $\mathcal{N}$ to the leading order. If $m_p = \mathcal{N}$, $\widetilde{\mathcal{N}} = 1$ and $p_f = 0$ in Eq.~\eqref{eq:pf}, as expected.

Note that if $n = m_p = 0$, $\widetilde{\mathcal{N}} =  \mathcal{N} + 1$ and $\gamma = 0$. Then $Z = \mathcal{N} + 1$ from Eq.~\eqref{eq:Z}, leading to the following simplification of
Eq.~\eqref{eq:pf}:
\begin{equation} \label{eq:pf:n0}
p_f = \frac{1}{\mathcal{N} (\mathcal{N} + 1)} \sum_{U^{i}=0}^{\mathcal{N}}  U^{i} = \frac{1}{2}.
\end{equation}
Thus, not surprisingly, the probability of finding a higher-fitness target before making any moves is $1/2$. After making a single move and not finding a higher-fitness target ($n=1$, $m_p=1$),
$\widetilde{\mathcal{N}} =  \mathcal{N}$ and $\gamma = 1/\mathcal{N}$. With the additional assumption that $\mathcal{N} \gg 1$, we obtain:
\begin{equation} \label{eq:pf:n1}
p_f \simeq \frac{1 - 2 e^{-1}}{1 - e^{-1}} + \mathcal{O} (\frac{1}{\mathcal{N}}) \simeq 0.42.
\end{equation}

In summary, the probability of finding a beneficial move, $p_f$, starts out at $0.5$ and decreases with the number of trials until either a beneficial move is found (in which case Eq.~\eqref{eq:pf} is no longer
applicable) or there are no more novel moves to find (in which case $p_f=0$). The asymptotic $\simeq 1/n$ behavior of $p_f$ is universal in the $n \gg 1$ limit (Eq.~\eqref{eq:pf:limit1}).

Finally, we observe that the above formalism can be extended to any subsets $S^i_p$ and $S^i$ since the initial division into deleterious/neutral moves in $S^i_p$ and beneficial moves
in $S^i$ was arbitrary. Thus, even if a beneficial move is found, we can add it to $S^i_p$ and regard $S^i$ as the set of \emph{remaining}, or \emph{novel} beneficial moves. \newline

\noindent
\textit{Weighted moves.} The probability to find a novel beneficial move (Eq.~\eqref{eq:pf}) was derived under the assumption that the total number of neighbors $\mathcal{N}$ is known and that the move set is unweighted -- each new move is chosen with equal probability $1/\mathcal{N}$. However, move sets may be intrinsically weighted: for example, in systems with recombination
relative weights of recombination moves depend on the genotype frequencies in the population. In addition, it may be of interest to assign separate weights to classes of moves,
such as one- and two-point mutations in sequence systems, or one- and two-spin flips in spin systems. In this section, we relax the assumption of unweighted moves,
while still treating $\mathcal{N}$ as a known constant. %\newline

Specifically, we consider a set of weights $\left\{ w_j \right\}_{j=1}^\mathcal{N}$ associated with $i \to j \in \text{nnb} (i)$ moves. The probability of a $i \to j$ jump is then given by
$p(i \to j) = {w_j}/{W}$, where $W = \sum_{j=1}^\mathcal{N} w_j = \sum_{j=1}^{U^i_p} w_j + \sum_{j=1}^{U^i} w_j \equiv W_{U^i_p} + W_{U^i}$ is the sum over all nearest-neighbor weights, and
$W_{U^i_p}$ and $W_{U^i}$ are partial sums over the weights in $S^i_p$ and $S^i$, respectively. Consequently,
\begin{equation} \label{eq:P:DU:w}
P(\mathcal{D} | U^{i}, \{ w_j \}) = \binom{n}{K} \left( \frac{W_{U^{i}}}{W} \right)^K \left( 1 - \frac{W_{U^{i}}}{W} \right)^{n-K},
\end{equation}
which in the $K = 0$ case reduces to
\begin{equation} \label{eq:P:DU:w:mod}
P(\mathcal{D} | U^{i}, \{ w_j \}) = \left( 1 - \frac{W_{U^{i}}}{W} \right)^{n} \simeq e^{-\frac{W_{U^{i}}}{W} n}.
\end{equation}
Next, we integrate the likelihood over the edge weights:
\begin{equation} \label{eq:P:DU:int}
P(\mathcal{D} | U^{i}) = \int_0^\infty dw_1 \dots dw_\mathcal{N} P(w_1) \dots P(w_\mathcal{N}) e^{-\frac{n}{W} (w_1 + \dots + w_{U^{i}})}.
\end{equation}
We represent the probability distribution of edge weights by a Gaussian mixture model, which can be used to describe multimodal distributions of arbitrary complexity~\cite{Bishop:2006}:
\begin{equation} \label{eq:Gaus:mix}
P(w) = \frac{1}{\Omega} \sum_{k=1}^\mathcal{P} \frac{p_k}{\sqrt{2 \pi} \sigma_k} e^{- \frac{(w - \bar{w}_k)^2}{2 \sigma_k^2}},
\end{equation}
where $\Omega$ is the normalization constant, $\mathcal{P}$ is the number of Gaussian components and $p_k$ is the relative weight of component $k$: $\sum_{k=1}^\mathcal{P} p_k = 1$.
In the $\mathcal{N} \gg 1$ limit, we expect $W \simeq \langle W \rangle = \mathcal{N} \sum_k p_k \bar{w}_k \equiv \mathcal{N} \bar{w}$, such that Eq.~\eqref{eq:P:DU:int} simplifies to
\begin{equation} \label{eq:P:DU:int:smpl}
P(\mathcal{D} | U^{i}) \simeq \prod_{j=1}^{U^{i}} \int_0^\infty dw_j P(w_j) e^{-\frac{w_j}{\langle W \rangle} n} = e^{-\beta U^{i}},
\end{equation}
where
\begin{equation} \label{eq:beta}
e^{-\beta} = \frac{1}{2 \Omega} \sum_{k=1}^\mathcal{P} p_k \text{erfc} \left( \frac{c_k - \bar{w}_k}{\sqrt{2} \sigma_k} \right) e^{-\alpha_k}.
\end{equation}
Here, $\alpha_k = \frac{n \bar{w}_k}{\langle W \rangle} - \frac{n^2 \sigma^2_k}{2 \langle W \rangle^2} = \gamma \frac{\bar{w}_k}{\bar{w}} - \gamma^2 \frac{\sigma^2_k}{2 \bar{w}^2}$,
$c_k = \frac{\sigma^2_k n}{\langle W \rangle} = \gamma \frac{\sigma^2_k}{\bar{w}}$ and $\text{erfc}(x) = \frac{2}{\sqrt{\pi}} \int_x^\infty dt e^{-t^2}$ is the complementary error function. The normalization constant is given by
\begin{equation} \label{eq:omega}
\Omega = \frac{1}{2} \sum_{k=1}^\mathcal{P} p_k \text{erfc} \left( -\frac{\bar{w}_k}{\sqrt{2} \sigma_k} \right).
\end{equation}
%$\alpha_k = \frac{n^2 \sigma^2_k}{2 \langle W \rangle^2} - \frac{n \bar{w}_k}{\langle W \rangle} + \log \frac{\sum_{k=1}^\mathcal{P} p_k \text{erfc} (-\bar{w}_k/\sqrt{2} \sigma_k)}{\text{erfc} ((c_k - \bar{w}_k)/\sqrt{2} \sigma_k)}$. Here $c_k = \sigma^2_k n/\langle W \rangle$ and $\text{erfc}(x)$ is the complementary error function.
Note that if all the Gaussians are narrow ($\sigma_k \ll \bar{w}_k$,  $\forall k$), $\text{erfc} \left( -\frac{\bar{w}_k}{\sqrt{2} \sigma_k} \right) \to 2$ and thus
$\Omega \to 1$, as expected.

If the edge weights are Gaussian distributed with mean $\bar{w}$ and standard deviation $\sigma$ (i.e., $\mathcal{P} = 1$), Eq.~\eqref{eq:beta} becomes
\begin{equation} \label{eq:beta:Gaus}
e^{-\beta} =  \frac{\text{erfc} \left( \frac{c - \bar{w}}{\sqrt{2} \sigma} \right)}{\text{erfc} \left( -\frac{\bar{w}}{\sqrt{2} \sigma} \right)} e^{-\alpha},
\end{equation}
where $\alpha = \gamma - \gamma^2 \frac{\sigma^2}{2 \bar{w}^2}$ and $c = \gamma \frac{\sigma^2}{\bar{w}}$. If in addition all weights are equal, $\frac{\sigma}{\bar{w}} \to 0$ and
$\beta \to \gamma$ in Eq.~\eqref{eq:beta:Gaus}, such that Eq.~\eqref{eq:P:DU:int:smpl} for the likelihood reduces to Eq.~\eqref{eq:P:DU:simple}. Thus, the difference between
$\beta$ and $\gamma$ is due to fluctuation corrections. The model evidence $Z$, the posterior probability $P(U^{i} | \mathcal{D})$ and $p_f$, the 
probability of finding a higher-fitness target in the next step, are found by substituting $\gamma \to \beta$ into Eqs.~\eqref{eq:Z}, \eqref{eq:P:UD2} and \eqref{eq:pf}, respectively.

Note that if $n \to 0$, $\beta \to 0$ as well and therefore $p_f \to 1/2$ since the argument leading to Eq.~\eqref{eq:pf:n0} still holds. Moreover, Eq.~\eqref{eq:pf} still yields
$p_f = 0$ when all the neighbors have been explored ($m_p = \mathcal{N}$). Finally, if $n, m_p \ll \mathcal{N}$ and $n \gg 1$, $\alpha \simeq \gamma$ and the ratio of
complementary error functions in Eq.~\eqref{eq:beta:Gaus} is $\simeq 1$. Then the argument leading to Eq.~\eqref{eq:pf:limit1} also holds, yielding $p_f \simeq 1/n$ asymptotically
even in the weighted move case. Thus, introducing weighted moves does not lead to qualitative differences in the $p_f$ dependence on $n$
%of the probability of finding a novel higher-fitness target on the number of trials $n$
-- any substantial differences are localized to the intermediate region: $1 \le n \le 30$ or so, and in many systems the $p_f$ curves for weighted
and unweighted moves overlap almost completely (cf. red and blue curves in Fig.~S1A-D).

Next, we consider the exponential probability distribution of edge weights -- an archetypical distribution often found in natural and artificial networks~\cite{Willow}:

\begin{equation} \label{eq:exp}
P(w) = \frac{1}{\bar{w}} e^{- \frac{w}{\bar{w}}},
\end{equation}
where $\bar{w}$ denotes the mean of the exponential distribution, such that $\langle W \rangle = \mathcal{N} \bar{w}$. It is easy to show that the likelihood $P(\mathcal{D} | U^{i})$ is given by Eq.~\eqref{eq:P:DU:int:smpl} with $\beta_\text{exp} = \log \left( 1 + \frac{n}{\mathcal{N}} \right)$. Consequently, as in the case of the Gaussian mixture model, the model evidence $Z$,
the posterior probability $P(U^{i} | \mathcal{D})$ and $p_f$ are given by Eqs.~\eqref{eq:Z}, \eqref{eq:P:UD2} and \eqref{eq:pf}, respectively, but with $\beta_\text{exp}$ instead of $\gamma$.
Clearly, $\beta_\text{exp} \to 0$ as $n \to 0$ and therefore $p_f \to 1/2$ as in the Gaussian mixture case. Similarly, $p_f = 0$ once $m_p =  \mathcal{N}$.
Lastly, the $n, m_p \ll \mathcal{N}$ limit yields $\alpha \simeq \gamma$, which in turn leads to $p_f \simeq 1/n$ under the additional assumption of $n \gg 1$.
Thus, the dependence of $p_f$ on $n$ for exponentially distributed weights is again qualitatively similar to the $p_f$ functions in the corresponding unweighted cases
(cf. red and blue curves in Fig.~S1E,F). \\

\noindent
\textit{Simplified treatment of $p_f$.} 
The computation of $p_f$ for the unweighted and the exponentially distributed cases requires the knowledge of $m_p$ and $\mathcal{N}$ besides the number of trials $n$.
For weighted move sets in the Gaussian mixture model, one would additionally require $p_k$, $\bar{w}_k$ and $\sigma_k$ for each Gaussian component.
Unless these parameters are known \textit{a priori}, they would have to be estimated from a sample of edge weights, increasing the computational burden.
Moreover, this extra effort may not be justified, in the view of the nearly universal dependence of $p_f$ on $n$ in all three cases considered above.
Even keeping track of $\mathcal{N}$ may be complicated for some move sets, e.g. a recombination+mutation move set employed in genetic algorithms~\cite{Goldberg:1989,Vikhar2016,Slowik:2020}.
With recombination, $\mathcal{N}$ depends on the current state of the population and therefore generally changes with time. 
Hence, computing $\mathcal{N}$ at each step would increase the complexity of the algorithm.
%Although this requirement is not too onerous, finding
%$\mathcal{N}$ may be difficult for complex move sets which depend on the current population state, such as recombination and mutation moves in genetic algorithms.

We propose to capitalize on the approximate universality of $p_f$ by creating a minimal model for it which depends only on the number of trials $n$.
Specifically, we define
\begin{equation} \label{pf:simple}
  p_f (n) =
    \begin{cases}
      \frac{n^2}{250} - \frac{2n}{25} + \frac{1}{2} & \text{if $0 \le n \le 5$}, \\
      \frac{1}{n} & \text{if $n > 5$}.
    \end{cases}
\end{equation}
This model has the right asymptotics at $n = 0$ and in the $n, m_p \ll \mathcal{N}$, $n \gg 1$ limit, but does not go to $0$ identically when $m_p = \mathcal{N}$ because enforcing this condition
requires the knowledge of $\mathcal{N}$. However, if $\mathcal{N} \gg 1$, as can be expected with complex move sets, $n \gg 1$ at $m_p = \mathcal{N}$ and the difference between
the small but non-zero value of $p_f$ in Eq.~\eqref{pf:simple} and zero will be immaterial (cf. green curves in Fig.~S1 for $p_f (n)$ in several model systems). \\

\noindent
\textbf{Implementation of  the optimal search policy: the SmartRunner algorithm.}
Given Bayesian probabilistic estimates of finding novel moves between a given node and its nearest neighbors, we need to formulate an optimal search policy in order to maximize the
expected fitness gain over the course of the run with $l_\mathrm{tot}$ random trials.

Assuming that the walker is currently at node $i$, there are two options after each random trial: stay at node $i$ (thereby rejecting the move) or jump to a neighboring node $j$. If the walker
stays at node $i$, we expect it to search for $l_i$ steps before finding a higher-fitness node which has not been detected before. Then the value of the policy of staying at $i$ can be
evaluated as
\begin{equation} \label{eq:stay}
\mathcal{S}_i = \xbar{\Delta \mathcal{F}_b} + \xbar{R}~\!(l_\mathrm{rem} - l_{i}) = - \xbar{R}~\!l_{i} + \mathcal{C},
\end{equation}
where $\xbar{\Delta \mathcal{F}_b} = \xbar{\mathcal{F}_k - \mathcal{F}_i}$ is the expected fitness gain of the newly found beneficial move to a node $k \in \text{nnb} (i)$
and $\xbar{R}$ is the expected rate of fitness gain per step times the number of steps remaining in the run.
%Note that the extra $-1$ accounts for the jump between the current node $i$ and the new beneficial node $k$.
Furthermore, $l_\mathrm{rem} \le l_\mathrm{tot}$ is the number of steps
remaining in the simulation, and $l_{i}$ is the expected number of steps needed to find $k$:
\begin{equation} \label{eq:l:xtra}
l_{i} = \mathrm{rnd}[\frac{1}{p^{i}_{f}}],
\end{equation}
where $p^{i}_{f}$ is given by Eq.~\eqref{eq:pf} or Eq.~\eqref{pf:simple} (with the dependence on the node index $i$ made explicit for clarity) and $\mathrm{rnd}[~]$ is the rounding operator.
Finally, $\mathcal{C} = \xbar{\Delta \mathcal{F}_b} + \xbar{R}~\!l_\mathrm{rem}$ denotes a constant contribution independent of the node index, under the assumption that
$\xbar{\Delta \mathcal{F}_b}$ is the same for all nodes.

The value of the policy of jumping to a downstream node $j$ from $i$ is given by
\begin{equation} \label{eq:leave}
\mathcal{L}_{i \to j} = (\mathcal{F}_{j} - \mathcal{F}_{i}) - \xbar{R}~\!(l_{j}+1) + \mathcal{C},
\end{equation}
where the extra factor of $-\xbar{R}$ accounts for the jump between the current node $i$ and the new node $j$, which reduces the total number of remaining steps by $1$.

\begin{figure}[!htb]
\centering
%\hspace*{-2.5cm}
\includegraphics[scale=0.55]{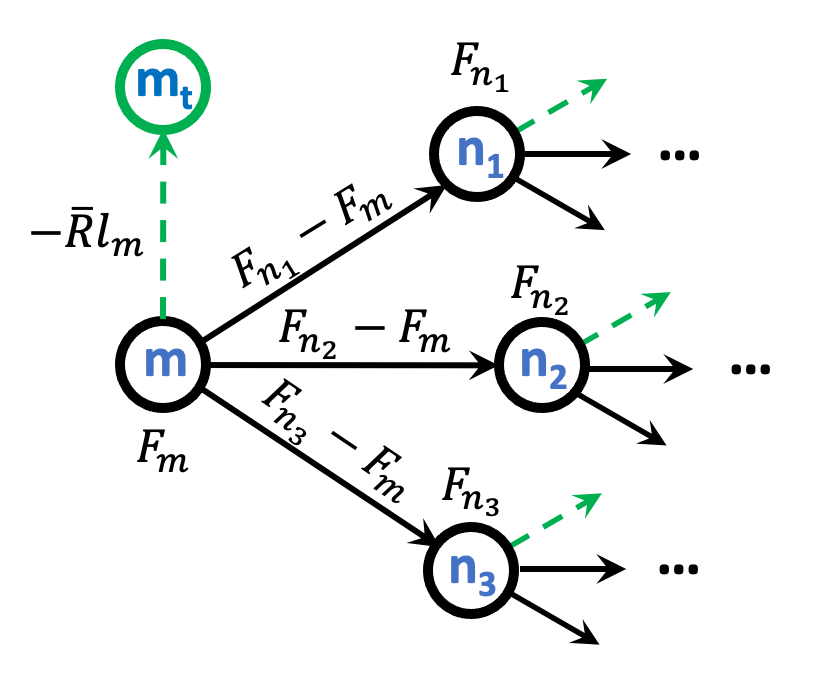}
\caption{
\textbf{A schematic representation of the directed graph $\mathcal{G}$ that represents the search process.}
Regular nodes (system states) are represented by black circles with the corresponding fitness values; terminal nodes are shown as green circles.
Directed edges connecting regular and terminal nodes are shown as dashed green lines and assigned a value of $-\xbar{R}~\!l_{m}$, where
$l_m$ is the expected number of steps to find a novel beneficial move starting from regular node $m$, and $\xbar{R}$ is the expected rate of fitness gain per step.
Directed edges connecting two regular nodes are shown as solid black lines and assigned a value of the fitness difference between the two nodes.
Note that the set of children of a given regular node always has one terminal node and $m_p = (0 \dots \mathcal{N})$ regular nodes depending on how much
exploration has been done. In general, $\mathcal{G}$ may contain directed cycles.
}
\label{fig:graph}
\end{figure}

We represent each visited state as a node and each rejected or accepted move as an edge on a directed
graph $\mathcal{G}$, implemented using the DiGraph class from NetworkX\footnote{\texttt{https://networkx.org/documentation/stable/reference/classes/digraph.html}}.
Thus, node $i$ is part of the directed graph $\mathcal{G}$ which contains information about all previously attempted moves.
Depending on which previous moves have been explored, node $i$ may be connected to multiple downstream nodes by directed edges; in general, these edges may form directed cycles.
To see if the jump to one of the nodes downstream of node $i$ will yield a more beneficial policy, we traverse $\mathcal{G}$ recursively starting from the node $i$, for up to
$l_\mathrm{max}$ steps. %Note that $l_\mathrm{max} = 1$ corresponds to stopping at the immediate downstream neighbors of the node $i$.
For computational convenience, $G$
is implemented with two types of nodes: `regular' nodes $i, j, k, \dots$ which denote states on the fitness landscape and are therefore assigned fitness values
$\mathcal{F}_{i}, \mathcal{F}_{j}, \mathcal{F}_{k}, \dots$ (black circles in Fig.~\ref{fig:graph}), and `terminal' nodes $i_t, j_t, k_t, \dots$ which are assigned fitness values
$-\xbar{R}~\!l_{i}, -\xbar{R}~\!l_{j}, -\xbar{R}~\!l_{k}, \dots$ (green circles in Fig.~\ref{fig:graph}).
Note that we have omitted the node-independent contribution $\mathcal{C}$ in Eqs.~\eqref{eq:stay} and \eqref{eq:leave}.
The edges of $\mathcal{G}$ connecting two regular nodes (solid black arrows in Fig.~\ref{fig:graph}): $m \to n$ are assigned a weight of $\mathcal{F}_{n} - \mathcal{F}_{m}$.
The edges of $\mathcal{G}$ connecting a regular node to a terminal node (dashed green arrows in Fig.~\ref{fig:graph}): $m \to m_t$ are assigned a weight of $- \xbar{R}~\!l_{m}$.
By construction, terminal nodes do not have descendants and each regular node has exactly one terminal descendant.
%\textcolor{red}{insert a graph Fig. here} 

The policy values are computed using a set of recursive paths on the directed graph $\mathcal{G}$.
All valid paths must start from node $i$ and end at one of the terminal nodes reachable with $\le l_\mathrm{max}$ steps.
The goal is to identify a path which has the maximum weight among all paths.
Note that with a single step, the only valid path is $i \to i_t$ and its weight is given by $\mathcal{S}_i$ from Eq.~\eqref{eq:stay}.
The minimum allowed value of $l_\mathrm{max}$ is thus equal to $2$ because this enables computations of the path weight as $\mathcal{L}_{i \to j}$ in Eq.~\eqref{eq:leave} for $j \in \text{nnb} (i)$.
Larger values of $l_\mathrm{max}$ will enable longer jumps if any are available; longer jumps entail repeated application of Eq.~\eqref{eq:leave} to compute the total path weight. 
%are still counted as a single move in Eqs.~\eqref{eq:stay} and \eqref{eq:leave};
%counting longer jumps as multiple moves would penalize them by lowering the number of remaining steps).
If the winning path is $i \to i_t$, the walker stays at the node $i$ and makes another random trial, updating its
$p^{i}_{f}$ accordingly. If the winning path is $i \to j_t$ (where $j$ may be several steps away depending on the value of $l_\mathrm{max}$), the walker jumps to the node $j$ and makes a new random trial
from that node. The node $j$ statistics such as $n$ and $m_p$ are initialized if the node has not been visited before in the run, and updated otherwise.

Note that if Eq.~\eqref{eq:pf} is used to compute $p^{i}_{f}$, it is possible to obtain $p^{i}_{f} = 0$ and therefore $l_{i} = \infty$ in Eq.~\eqref{eq:l:xtra},
which is represented computationally by a large positive constant. The case in which both node $i$ and all its neighbors $j$ reachable in $\le l_\mathrm{max}$ steps are in this category
requires special treatment because the $\xbar{R}~\!l_{i}$ and $\xbar{R}~\!l_{j}$ penalties cancel out and SmartRunner essentially becomes a local optimizer driven solely by fitness differences.
To avoid being trapped in local fitness maxima in this special case, SmartRunner employs two alternative strategies. In the first strategy, a random path is chosen in the ensemble of all paths with $\le l_\mathrm{max}$ steps, instead of the path with the maximum sum of edge weights. In the second strategy, a longer random path to the boundary of the $p_f = 0$ region is constructed explicitly; the random path can have up to $10^3$ steps. In both strategies, if the boundary of the $p_f = 0$ region is not reached, the procedure is repeated at subsequent steps, resulting in an overall random walk to the boundary of the
``maximally-explored'' region.

The SmartRunner algorithm depends on $\xbar{R}$, the expected rate of fitness gain per step. Larger positive values of $\xbar{R}$ will encourage jumps to less-explored regular nodes even if those have
slightly lower fitness values and will therefore promote landscape exploration. Smaller positive values of $\xbar{R}$ will encourage more thorough exploration of the current node but will not fully
prevent deleterious moves. Negative values of $\xbar{R}$ however will prevent all further exploration.
We adjust the value of $\xbar{R}$ adaptively as follows. The algorithm starts out with a user-provided initial value $\xbar{R}_\mathrm{~\!\!init}$. For each move, either accepted or rejected, the corresponding
fitness value is recorded in a fitness trajectory array. Once $M$ values are accumulated in the array, a linear model is fit to the fitness trajectory, yielding the fitted slope $R^\mathrm{fit}$.
Finally, $\xbar{R}$ is computed as
\begin{equation} \label{eq:Rbar}
\xbar{R} = 
    \begin{cases}
      \alpha R^\mathrm{fit} & \text{if $R^\mathrm{fit} \ge \epsilon$}, \\
      \alpha \epsilon \exp({R^\mathrm{fit} - \epsilon}) & \text{if $R^\mathrm{fit} < \epsilon$}, 
    \end{cases}
\end{equation}
where $\epsilon$ is a small positive constant. Note that the second line serves to `rectify' the values of $R^\mathrm{fit}$ that follow below the threshold $\epsilon$, preventing
$\xbar{R}$ from ever reaching negative values. The value of $\xbar{R}$ is recomputed every $M$ steps using Eq.~\eqref{eq:Rbar}, providing adaptive feedback throughout the run.
The positive hyperparameter $\alpha$ is the level of `optimism' -- how much more optimistic the system is about its future success compared to past performance. As discussed above,
larger values of $\alpha$ will promote landscape exploration. \newline

\noindent
The SmartRunner algorithm can be summarized as the following sequence of steps:
%\begin{enumerate}
%\item $U'\leftarrow U$
%\item Use random-walk propagator on $U$ to compute the matrix $V$ (equation \ref{eq:3}):\\ $V_{nc}\leftarrow\sum_{l=1}^{l_{max}}\left(\sum_{n'}(A^{l})_{nn'}w_{n'}U_{n'c}\right)$
%\item Compute the rate of transition between communities:\\ $Q_{cc'}\leftarrow\frac{(V^{T}U)_{cc'}}{\sum_{n}w_{n}U_{nc'}}$
%\item Based on posterior probabilities, for each node $n$ reassign $U_{nc}\leftarrow\delta_{\tilde{c}_{n}c}$ (equation \ref{eq:5} and \ref{eq:6} with $g_c=\W_c/Q_{cc}$) for \\ $\tilde{c}_{n}=\text{argmax}_{c''}\left(\sum_{c'}\frac{1}{Q_{c'c'}}(V_{nc'}\log(Q_{c'c''})-Q_{c'c''}w_{n})\right)$
%\end{enumerate}
%\pagebreak
\vspace{20pt}

\hrule
\hrule
\hrule
\vspace{5pt}
\noindent
%\textbf{FUNCTION}: Walk-likelihood function
\textbf{SmartRunner Algorithm}
\vspace{5pt}
\hrule
\hrule
\hrule
\vspace{5pt}
\noindent
\textbf{INPUT:} \\
Initial state: $X_0$\\
Fitness landscape function: $X \to \mathcal{F}$\\
Move set function: $X^\mathrm{old} \to X^\mathrm{new}$\\
Total number of iterations: $l_\mathrm{tot}$\\
Maximum length of paths explored from each state: $l_\mathrm{max}$\\
Initial guess of the fitness rate: $\xbar{R}_0$\\
Optimism level: $\alpha$ \\
Length of sub-trajectory for recomputing $\xbar{R}$: $M$
\vspace{5pt}
\hrule
\vspace{5pt}
\begin{enumerate}
\item Initialize directed graph $\mathcal{G}$.
\item Initialize regular node $X_0$ with $\mathcal{F} (X_0)$.
\item Initialize terminal node $X_{0,t}$.
\item Initialize $\xbar{R} = \xbar{R}_0$.
\item Initialize $l = 0$.
\item Add an edge $X_0 \to X_{0,t}$ with a weight $-\xbar{R}~\!l_{X_0}$.
\end{enumerate}
\noindent
\textbf{do:}
\begin{enumerate}
\item Generate a random move: $X \to X'$. %\hfill
\item If $X' \notin \mathcal{G}$: add $X'$ to $\mathcal{G}$ with $\mathcal{F} (X')$; add a terminal node $X'_t$; add an edge $X' \to X'_{t}$ with a weight $-\xbar{R}~\!l_{X'}$.
\item If $X \to X' \notin \mathcal{G}$: add an edge $X \to X'$ with a weight $\mathcal{F} (X') - \mathcal{F} (X)$.
\item Update statistics for $X$, recompute $l_{X}$ and update the $X \to X_t$ edge weight.
\item Recursively compute sums of edge weights for all paths of length $\le l_\mathrm{max}$ starting at $X$ and ending at downstream terminal nodes. If $l_X = \infty$ for the $X X_t$ path and
$l_{Y_k} = \infty$ for all other paths $X \dots Y_k Y_{k,t}$ in the ensemble, initiate a random walk; otherwise, stay at $X$ or jump to $Y_k$ according to the path with the maximum sum of edge weights.
\item If $l = M, 2M, 3M, \dots$: recompute $\xbar{R}$ using Eq.~\eqref{eq:Rbar}.
%\item $Q_{cc'} \leftarrow (V^{T} U')_{cc'} / \sum_{n=1}^N w_{n} {U'}_{nc'}$ \hfill Eq.~\eqref{eq:Q}
\end{enumerate}
\noindent
\textbf{while} $l \le l_\mathrm{tot}$
\vspace{5pt}
\hrule
\vspace{5pt}
\noindent
\textbf{OUTPUT:} \\
Globally best state: $X^\mathrm{best}$, $\mathcal{F} (X^\mathrm{best})$\\
Fitness trajectory: $\{ \mathcal{F} \}_{l=1}^{l_\mathrm{tot}}$\\
Total number of fitness function evaluations: $f_\mathrm{eval}$
\vspace{5pt}
\hrule

\vspace{1cm}
\noindent
\textbf{Adaptive fitness landscape.}
The stay or leave policy defined by Eqs.~\eqref{eq:stay} and \eqref{eq:leave} amounts to an adaptive redefinition of the fitness landscape:
\begin{equation} \label{Ftilde}
\mathcal{F}_{i} \to \widetilde{\mathcal{F}}_{i} = \mathcal{F}_{i} - \xbar{R}~\!l_{i},
\end{equation}
where $\xbar{R}~\!l_{i}$ is a positive occupancy penalty whose overall magnitude is controlled by the hyperparameter $\xbar{R}$. The penalty increases as the node $i$ is explored more and more,
resulting in progressively larger values of $l_i$. Note that if Eq.~\eqref{pf:simple} is used to estimate $p_f$, the only additional piece of information required to compute $\widetilde{\mathcal{F}}_{i}$
from $\mathcal{F}_{i}$ is the total number of trials $n_i$ at the node $i$, which is easy to keep track of.
Thus, $\widetilde{\mathcal{F}}_{i}$ can serve as input not only to SmartRunner, which in this view amounts to hill climbing on the $\widetilde{\mathcal{F}}$
landscape, but to any global optimization algorithm. % with an arbitrary move set.
In algorithms where individual moves are accepted or rejected sequentially (e.g., Simulated Annealing, Stochastic Hill Climbing), we compare $\widetilde{\mathcal{F}}_{i}$ with
$\widetilde{\mathcal{F}}_{j} - \xbar{R}$ to account for the fact that jumping from node $i$ to node $j$ decreases the total number of remaining steps by $1$ (cf. Eq.~\eqref{eq:leave}).
In algorithms which involve non-sequential scenarios (e.g, Evolutionary Algorithm), modified fitnesses $\widetilde{\mathcal{F}}$ from Eq.~\eqref{Ftilde} are used directly instead of $\mathcal{F}$.

\section*{Results}

\begin{figure}[!htb]
\centering
%\hspace*{-2.5cm}
\includegraphics[scale=0.55]{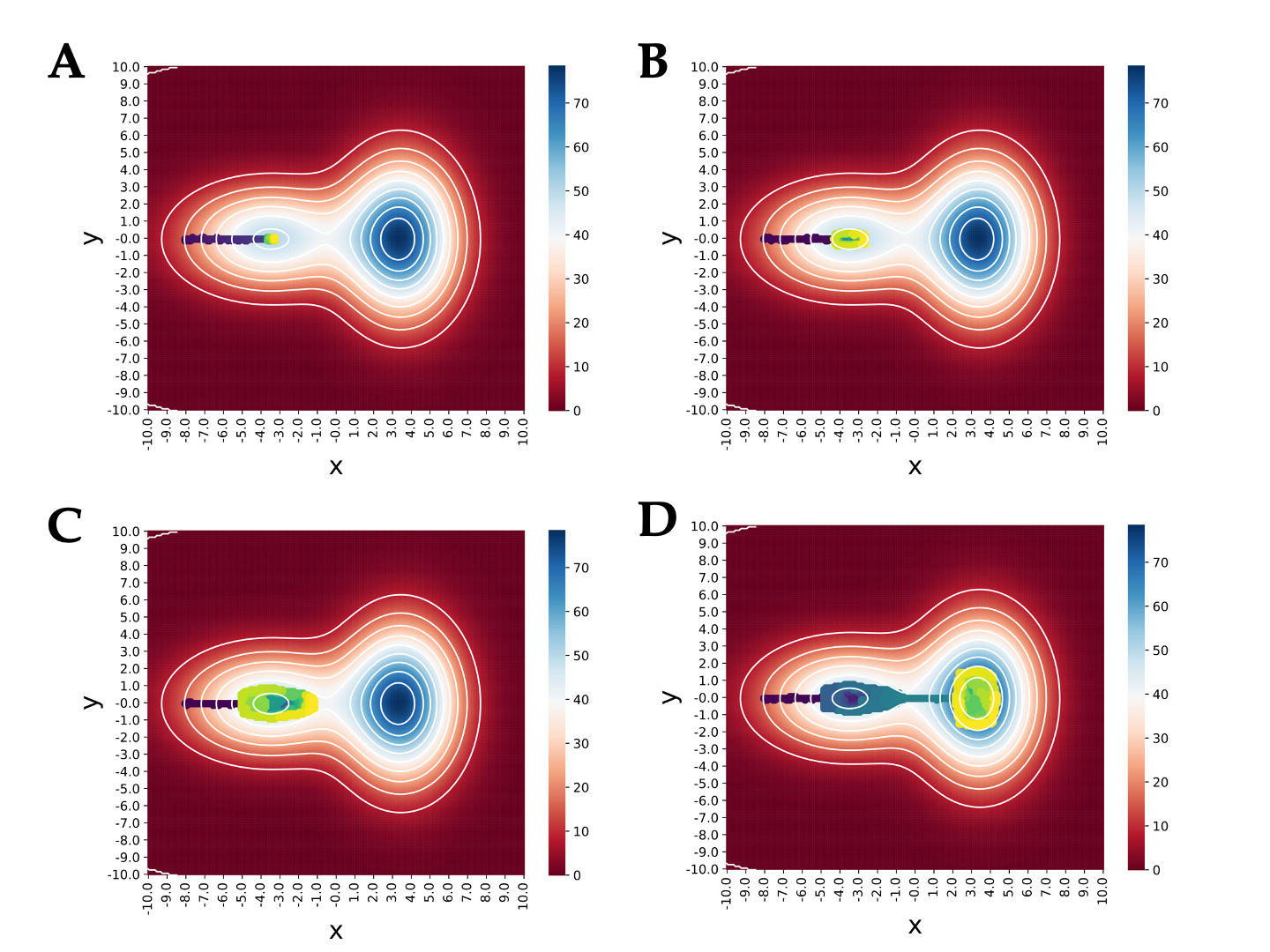}
\caption{
\textbf{SmartRunner search for the global maximum on a 2D fitness landscape with two basins of attraction.}
The fitness function is given by a weighted sum of two Gaussians:
$\mathcal{F} (x,y) = \sum_{k=1}^{2} A_k \exp \left[ -{(x - \bar{x}_k)^2}/{2 \sigma_{x,k}^2} - {(y - \bar{y}_k)^2}/{2 \sigma_{y,k}^2} \right]$, with
$(A_1, A_2) = (50, 75)$, $(\bar{x}_1, \bar{y}_1) = (-3.5, 0.0)$, $(\sigma_{x,1}, \sigma_{y,1}) = (3.0, 2.0)$,
$(\bar{x}_2, \bar{y}_2) = (3.5, 0.0)$, and $(\sigma_{x,2}, \sigma_{y,2}) = (2.0, 3.0)$.
The fitness landscape is discretized with a $0.01$ step in both $x$- and $y$-directions, with $x \in [-10,10]$, $y \in [-10,10]$ and
periodic boundary conditions; the total number of states is $4 \times 10^6$.
Panels A-D show $4$ SmartRunner trajectories starting from $(x_0, y_0) = (-8.0, 0.0)$, with $l_\mathrm{tot} = 10^4, 10^5, 5 \times 10^5, \text{ and } 10^6$,
respectively. The trajectories are color-coded from dark blue to yellow as the run progresses. All runs used $\alpha = 0.1$, $l_\mathrm{max} = 2$,  and $\xbar{R}^\mathrm{init} = 0.1$.
The best fitness value found in the run is $\mathcal{F}_\mathrm{best} = 50.17$ (left peak) in panels A-C and $78.48$ (right peak) in panel D.
}
\label{fig:2Dsearch}
\end{figure}

\noindent
\textbf{SmartRunner can climb out of deep local maxima.} To demonstrate the ability of SmartRunner to traverse local basins of attraction leading to suboptimal solutions, we have constructed a
2D fitness landscape defined by a weighted sum of two Gaussians (Fig.~\ref{fig:2Dsearch}). The left basin of attraction leads to a local maximum ($\mathcal{F}^\star = 50.17$) which is much
smaller compared to the global maximum on the right ($\mathcal{F}^\star = 78.48$). The two basins of attraction are separated by a steep barrier. We start the SmartRunner runs from the left
of the local basin of attraction, making sure that the walker rapidly falls there first, reaching the local maximum in a few thousand steps (Fig.~\ref{fig:2Dsearch}A). Afterwards, the walker explores
the local basin of attraction more and more extensively (Fig.~\ref{fig:2Dsearch}B,C) until the barrier is finally overcome and the global maximum is found (Fig.~\ref{fig:2Dsearch}D).
The exploration strategy is automatically adapted to the fitness landscape features rather than being driven by external parameters such as the simulated annealing temperature.

%%%%%%

\vspace{0.4cm}
\noindent
\textbf{SmartRunner performance on 4D test functions.} Next, we have explored SmartRunner performance on three standard 4D test functions often used to benchmark
global optimization algorithms~\cite{Torn:1989}: Rastrigin, Ackley and Griewank (se SI Methods for function definitions). The test functions are defined in standard hypercube ranges
and supplemented with periodic boundary conditions. The resulting fitness landscapes are discretized using the same step size $\Delta x$ in all $4$ directions, resulting in $1.63 \times 10^9$,
$1.17 \times 10^{10}$ and $2.08 \times 10^{12}$ distinct fitness states for Rastrigin, Ackley and Griewank functions, respectively. All three test functions are characterized by
multiple local maxima; the unique global maximum is located at $\vec{\bf x} = (0,0,0,0)$ and corresponds to $\mathcal{F} = 0$. The landscapes are explored by randomly choosing
one of the directions and then increasing or decreasing the corresponding coordinate by $\Delta x$ (the \textit{nnb} moveset).

\begin{figure}[!htb]
\centering
%\hspace*{-2.5cm}
\includegraphics[scale=0.60]{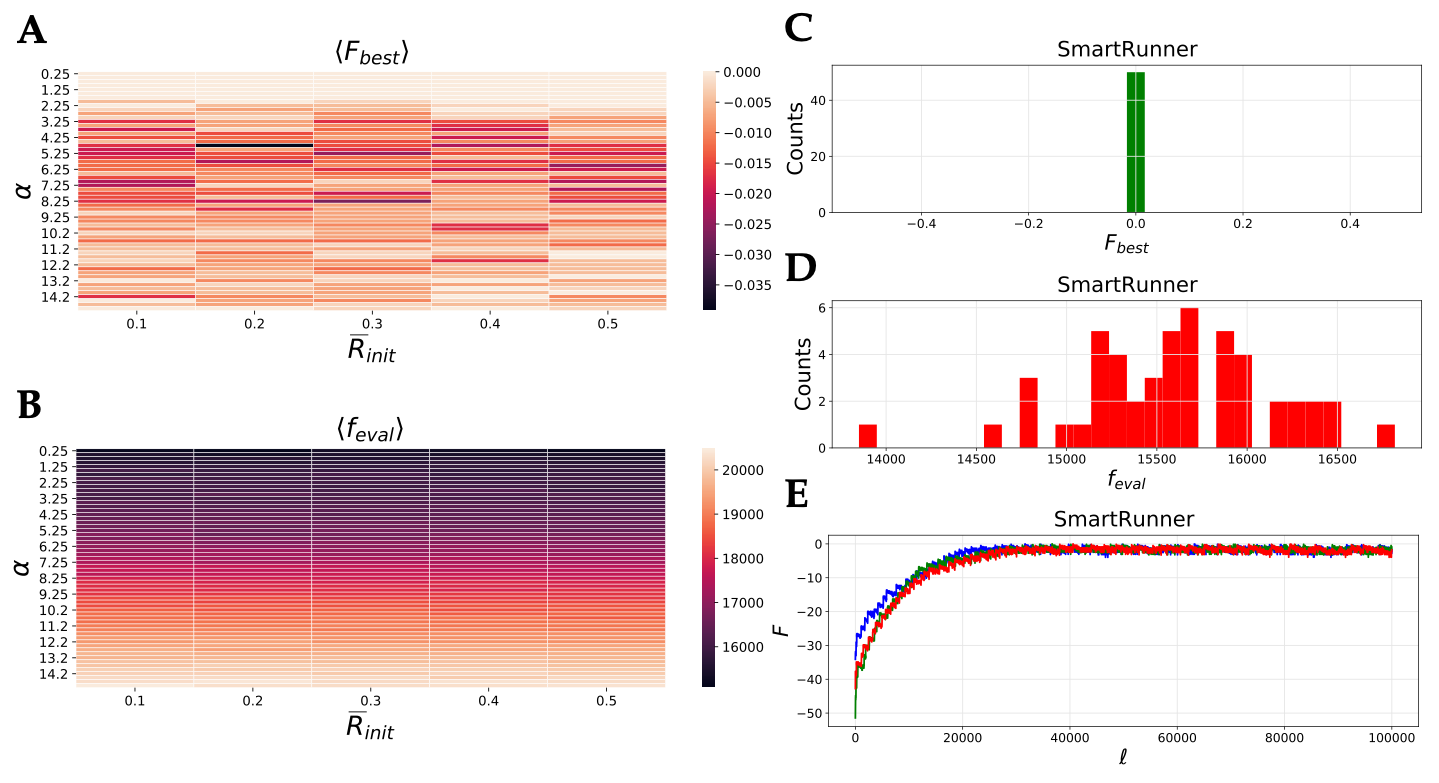}
\caption{
\textbf{SmartRunner exploration of the Rastrigin test function: \textit{nnb} moveset.}
(A) A scan over SmartRunner hyperparameters ($l_\mathrm{max} = 2$): the initial value of the expected rate of fitness gain per step $\xbar{R}_\mathrm{~\!\!init}$ and the level of optimism $\alpha$.
Each cell in the heatmap represents best fitness values found in each run, averaged over $50$ independent runs with $l_\mathrm{tot} = 10^5$ steps each and randomly chosen starting states.
(B) Same as A but with the average taken over the number of fitness function evaluations (unique fitness function calls) in each run.
(C) A histogram of best fitness values for the heatmap cell with $\xbar{R}_\mathrm{~\!\!init} = 0.1$, $\alpha = 1.0$.
(D) A histogram of the number of unique fitness function calls for the heatmap cell with $\xbar{R}_\mathrm{~\!\!init} = 0.1$, $\alpha = 1.0$.
(E) Plots of 3 representative SmartRunner trajectories ($\xbar{R}_\mathrm{~\!\!init} = 0.1$, $\alpha = 1.0$).
}
\label{fig:SRruns:R}
\end{figure}

Fig.~\ref{fig:SRruns:R} shows the performance of SmartRunner on the Rastrigin test function: Fig.~\ref{fig:SRruns:R}A is a hyperparameter scan which shows no consistent trend
in the dependence of the average best fitness values $\langle \mathcal{F}_\mathrm{best} \rangle$ on $\xbar{R}_\mathrm{~\!\!init}$, the initial rate of fitness gain per step. This is expected
because the value of $\xbar{R}$ is reset adaptively during the run (cf. Eq.~\eqref{eq:Rbar}). In contrast, there is a slight preference for lower values of optimism $\alpha$.
Fig.~\ref{fig:SRruns:R}B shows the corresponding average of function evaluations -- unique fitness function calls which can be used as a measure of algorithm performance,
especially in cases where fitness function calls are expensive, making it advisable to focus on maximizing the average fitness gain per function evaluation. As expected, the optimal values of $\alpha$
correspond to the lower number of function evaluations since lower values of $\alpha$ tend to favor exploitation (i..e., a more thorough search of the neighbors of the current state) over exploration
(which favors more frequent jumps between landscape states). Figs.~S2A,B and S2C,D show the corresponding results for Ackley and Griewank test functions, respectively.
Lower values of $\alpha$ work better for Ackley, while $\alpha \geq 5$ are preferable for Griewank, indicating that in general a scan over several values of $\alpha$ may be required.
Since the Griewank landscape is considerably larger and the global maximum is not always found, we also show the maximum best-fitness value found over $50$ independent runs, and the
corresponding number of function evaluations (Fig.~S2E,F). For lower values of $\alpha$, the global maximum is not always found but rather another high-fitness solution.
With reasonable hyperparameter settings, all $50$ SmartRunner runs find the global maximum of the Rastrigin landscape (Fig.~\ref{fig:SRruns:R}C), requiring $\simeq 15500$ function evaluations on
average. Fig.~\ref{fig:SRruns:R}E shows three representative fitness trajectories -- rapid convergence to the vicinity of the global maximum is observed in $\le 4 \times 10^4$ steps, regardless of the starting state.

\begin{figure}[!htb]
\centering
%\hspace*{-2.5cm}
\includegraphics[scale=0.60]{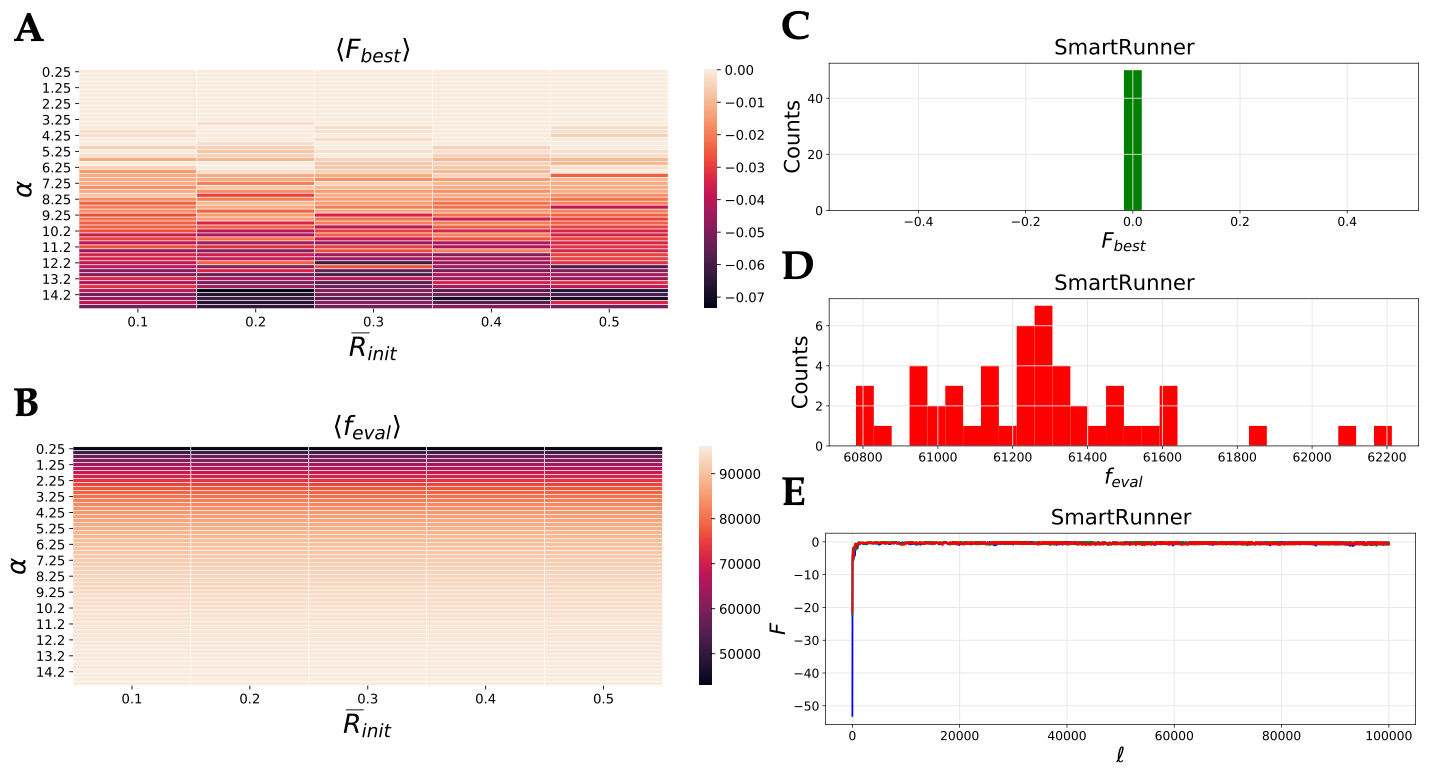}
\caption{
\textbf{SmartRunner exploration of the Rastrigin test function: \textit{spmut} moveset.}
Same as Fig.~\ref{fig:SRruns:R} (including SmartRunner settings and hyperparameter value settings in panels C-E), but with the $spmut$ moveset.
}
\label{fig:SRruns:Rspmut}
\end{figure}

The dynamics of global optimization strongly depend on the moveset type. To explore whether SmartRunner can adapt to movesets with non-local moves, we have considered the \textit{spmut} moveset in 
which a randomly chosen coordinate is changed to an arbitrary new value on the discretized landscape. Thus, instead of updating a given coordinate in $\pm \Delta x$ increments, most moves
change the coordinate by many multiples of $\Delta x$, creating a densely connected landscape: for example, the number of nearest neighbors is $200 \times 4 = 800$ for the
Rastrigin function, instead of just $8$ with the \textit{nnb} moveset (the abbreviation \textit{spmut} stands for single-point mutations,
since a given $x_i$ can `mutate' into any other $x_j$, $j \ne i$ from a discrete set). Fig.~\ref{fig:SRruns:Rspmut} shows that SmartRunner reliably finds the global maximum with the \textit{spmut}
moveset. The dependence on $\xbar{R}_\mathrm{~\!\!init}$ is weak and the lower values of $\alpha$ are preferable (Fig.~\ref{fig:SRruns:Rspmut}A). The number of fitness function calls is much higher
for the same total number of steps ($10^5$) as with the \textit{nnb} moveset (Fig.~\ref{fig:SRruns:Rspmut}B,D).
% , consistent with the much larger number of nearest neighbors
All $50$ runs find the global maximum with optimal or nearly-optimal hyperparameter settings (Fig.~\ref{fig:SRruns:Rspmut}C), and fitness trajectories quickly converge to high-quality
solutions (Fig.~\ref{fig:SRruns:Rspmut}E). Similar behavior is observed with Ackley and Griewank functions: lower values of $\alpha$ work better and the number of function evaluations is several
times larger compared to the \textit{nnb} moveset (Fig.~S3). Thus, using the \textit{nnb} moveset is preferable for all three landscapes.

%%%%%%

Next, we have asked whether the observed differences in SmartRunner performance at different hyperparameter values are statistically significant. Using the Rastrigin function as an example, we have
employed one-sided Kolmogorov-Smirnov (KS) tests for the best-fitness distributions (Fig.~S4). The distribution with the highest average of best-fitness values in Figs.~\ref{fig:SRruns:R}A and
\ref{fig:SRruns:Rspmut}A was compared with all the other distributions. We find that the differences between the distributions are not statistically significant with the \textit{nnb} moveset (Fig.~S4A).
In contrast, using $\alpha \ge 6$ with the \textit{spmut} moveset leads to statistically significant degradation of SmartRunner performance (Fig.~S4B).
We have also used KS tests to investigate the effects of the $l_\mathrm{max}$ hyperparameter (Fig.~S5). Since for all three test functions best-fitness distributions with $l_\mathrm{max} = 3$
are not significantly better than the corresponding $l_\mathrm{max} = 2$ distributions with the same $\alpha$ and $\xbar{R}_\mathrm{~\!\!init}$ hyperparameter settings, we typically use
$l_\mathrm{max} = 2$ as it is less expensive computationally.

\vspace{0.4cm}
\noindent
\textbf{The effects of the occupancy penalty on other global optimization algorithms.}  As mentioned above, the SmartRunner algorithm can be viewed as hill climbing on a fitness landscape $\widetilde{\mathcal{F}}$
modified with the occupancy penalties (Eq.~\eqref{Ftilde}). However, the modified fitness landscape can also be explored using other empirical global optimization approaches. Here, we focus on three
widely used algorithms: Simulated Annealing (SA)~\cite{Kirkpatrick:1983}, Stochastic Hill Climbing (SHC)~\cite{Juels1995}, and Evolutionary Algorithm (EA)~\cite{Goldberg:1989,Vikhar2016,Slowik:2020}
(see SI Methods for implementation details). SA is based on an analogy with
a metallurgy technique involving heating followed by controlled cooling of a material to alter its physical properties~\cite{Kirkpatrick:1983}. The algorithm is implemented as a series of Metropolis
Monte Carlo move trials~\cite{Metropolis:1953} with a slowly decreasing temperature. SA's hyperparameters are the initial temperature $T_i$ and the final temperature $T_f$, plus the expected rate
of fitness gain $\xbar{R}$ when the occupancy penalty is included. We use a linear cooling schedule in this work. SHC is a version of hill climbing which accepts downhill moves with the probability
$p = 1/(1 + \exp{\left[ (\mathcal{F}_\text{current} - \mathcal{F}_\text{new})/T \right]} )$~\cite{Juels1995}. Thus, $p \simeq 0$ in the $\mathcal{F}_\text{new}/T \ll \mathcal{F}_\text{current}/T$ limit,
and $p \simeq 1$ in the opposite limit. SHC's search strategy is controlled by the temperature $T$, along with $\xbar{R}$ in the case of modified landscapes.
Finally, EA is inspired by the process of biological evolution~\cite{Vikhar2016,Slowik:2020}. It involves creating a population of $N_\text{pop}$ `organisms'
(i.e., putative solutions; we use $N_\text{pop} = 50$ in this work). The population is initialized randomly and subjected to repeated rounds of recombination, mutation and selection.
Besides the population size, EA's hyperparameters are the crossover (recombination) rate $r_x$,
the mutation rate $\mu$ and, for modified landscapes, the expected rate of fitness gain $\xbar{R}$.

The original algorithm names (SA, SHC, EA) are reserved for runs with $\xbar{R} = 0$; runs with modified landscapes are referred to as `enhanced' (ESA, ESHC, EEA).
Fig.~S6 shows the performance of ESA as a function of the initial temperature $T_i$ and the expected rate of fitness gain $\xbar{R}$ for our three test functions, with the \textit{nnb} moveset
(although we have also performed a scan over the final temperature $T_f$, the dependence is weak and the results are not shown).
We observe that $T_i \simeq 1$ values are more preferable and, as expected, are accompanied by the lower number of function evaluations. Strikingly,
the hyperparameter settings with the best average performance always have non-zero $\xbar{R}$~\!: $T_i = 1.0$, $T_f = 0.002$, $\xbar{R} = 0.1$ for the Rastrigin function
(the corresponding $\langle \mathcal{F}_\mathrm{best} \rangle = -0.017$). For the Ackley function, $T_i = 1.0$, $T_f = 0.001$, $\xbar{R} = 0.15$
(the corresponding $\langle \mathcal{F}_\mathrm{best} \rangle = -2.078$). For the Griewank function, $T_i = 1.0$, $T_f = 0.001$, $\xbar{R} = 0.2$
(the corresponding $\langle \mathcal{F}_\mathrm{best} \rangle = -0.067$). Thus, ESA outperforms SA -- when using simulated annealing, the best global optimization strategy is to augment
the original fitness values with the occupancy penalties. Fig.~S7 shows that these observations are statistically significant. %, especially for the Rastrigin and Griewank functions.

%%%%%%
\begin{figure}[!htb]
\centering
%\hspace*{-2.5cm}
\includegraphics[scale=0.60]{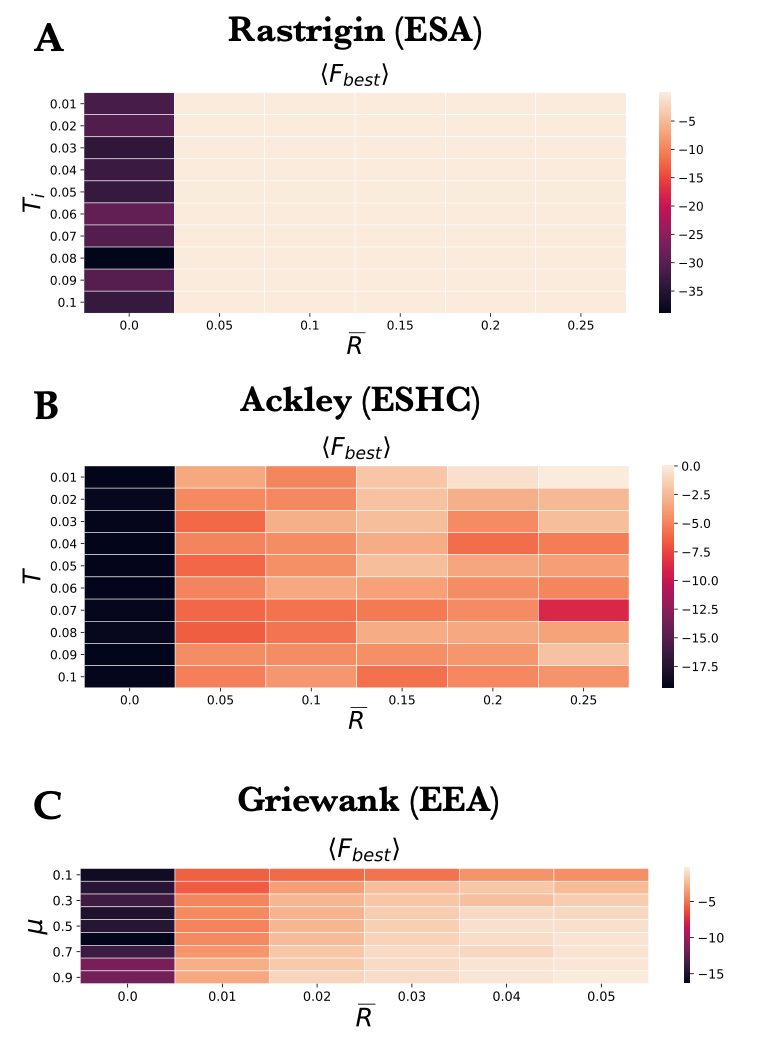}
\caption{
\textbf{Occupancy penalties enhance performance of global optimization algorithms: \textit{nnb} moveset.}
(A) A scan over ESA hyperparameters (linear cooling schedule) for the Rastrigin function:
the expected rate of fitness gain per step $\xbar{R}$ and the initial temperature $T_i$ (the final temperature is set to $T_f = 0.001$).
See Fig.~S8 for details and for the Ackley and Griewank functions.
(B) A scan over ESHC hyperparameters for the Ackley function:
the expected rate of fitness gain per step $\xbar{R}$ and the temperature $T$.
See Fig.~S11 for details and for the Rastrigin and Griewank functions.
(C) A scan over EEA hyperparameters for the Griewank function:
the expected rate of fitness gain per step $\xbar{R}$ and the mutation rate $\mu$ (the crossover rate is set to $r_x = 0.2$ and the population size to $N_\text{pop} = 50$).
See Fig.~S12 for details and for the Rastrigin and Ackley functions.
Each heatmap cell represents best fitness values found in each run, averaged over $50$ independent runs with $l_\mathrm{tot} = 10^5$ steps each and randomly chosen starting states.
}
\label{fig:tunnel}
\end{figure}

Occupancy penalties dramatically improve SA's performance when it is run with the suboptimal values of the initial temperature (Fig.~\ref{fig:tunnel}A, Fig.~S8). In fact, the results are better than those with the higher,
SA-optimal values of $T_i$: $\langle \mathcal{F}_\mathrm{best} \rangle = -0.005$ for Rastrigin ($T_i = 0.02$, $T_f = 0.003$, $\xbar{R} = 0.2$),
$\langle \mathcal{F}_\mathrm{best} \rangle = 0.000$ for Ackley ($T_i = 0.01$, $T_f = 0.001$, $\xbar{R} = 0.25$),
$\langle \mathcal{F}_\mathrm{best} \rangle = -0.015$ for Griewank ($T_i = 0.01$, $T_f = 0.003$, $\xbar{R} = 0.25$).
Thus, the best overall strategy is to run SA at very low temperatures (where it reduces to simple hill climbing), but on the modified fitness landscape. This is precisely the strategy implemented in
SmartRunner.

Qualitatively similar results are obtained with SHC: non-zero values of $\xbar{R}$ are preferable at higher, SHC-optimal values of $T$ (Fig.~S9); the effect is statistically significant (Fig.~S10).
However, as Fig.~\ref{fig:tunnel}B and Fig.~S11 demonstrate, the enhancement is especially dramatic when the values of $T$ become very low, much lower than the SHC-optimal values explored in Fig.~S9.
Similar to SA, low-$T$ runs with $\xbar{R} \ne 0$ yield the highest-quality solutions, again indicating that in the presence of occupancy penalties the best strategy is straightforward hill ascent.

Finally, occupancy penalties can rescue EA from being stuck in the local maxima (Fig.~\ref{fig:tunnel}C, Fig.~S12A,C) -- with the \textit{nnb} moveset, the population tends to condense onto a local maximum and
become monomorphic. Local mutations of population members in such locally optimal states are mostly deleterious and therefore tend to get eliminated from the population. The population as
a whole is therefore unable to keep exploring new states, as evidenced by the low number of function evaluations in Fig.~S12B,D,F compared to the other algorithms. This drawback is fixed
by making the fitness landscape adaptive with the help of the occupancy penalty.

SmartRunner can be viewed as stochastic generalization of the Taboo Search (TS) -- a deterministic policy in which all nearest neighbors of the current state are explored one by one and the
move to the neighbor state with the best fitness is accepted~\cite{Cvijovic:1995}. To prevent backtracking to already-explored states, a list of `taboo' states is kept to which jumps are forbidden;
the length of this list, $L_\text{tabu}$, is a hyperparameter. By construction, TS avoids visiting neighbor states more than once and is always guaranteed to find the best neighboring state to jump into;
however, we expect it to lose efficiency in systems characterized by very large numbers of neighbors, since all of these neighbors have to be tried and most of them do not correspond to good solutions.
In contrast, SmartRunner can make a decision to accept a move before all neighbors are explored, based on the move/neighbor statistics collected up to that point.
In any event, with $L_\text{tabu} \ge 400$ TS demonstrates high performance on all three test functions, requiring a relatively low number of function evaluations to achieve this result (Fig.~S13).

Interestingly, the situation is reversed with the \textit{spmut} moveset -- with SA and SHC, better performance is achieved when $\xbar{R} = 0$ (Fig.~S14). This observation is not surprising given the somewhat
special nature of the test functions we consider. As it turns out, with Rastrigin and Ackley functions it is possible to use TS to reach the global maximum in exactly $4$ steps, regardless of the initial
state. Each step sets one of the coordinates to $0.0$ until the global maximum is attained (see Fig.~S15A,B for representative trajectories). With the Griewank function, optimization
depends on the initial conditions and, as a rule, additional steps are required since the first $4$ steps only bring the system to the vicinity of the global maximum (Fig.~S15C).
Thus, with this landscape structure it is not beneficial to jump to a new node before all neighbors of the current node are explored. %-- all nodes are equivalent or nearly equivalent, and
In other words, premature jumping between nodes simply resets the search. In this case, $\xbar{R} = 0$ is indeed preferable and correctly identified by our methods;
however, this is a special case which we do not expect to hold true in general.
%the large number of neighbors created by the \textit{spmut} moveset

\begin{figure}[!htb]
\centering
%\hspace*{-2.5cm}
\includegraphics[scale=0.65]{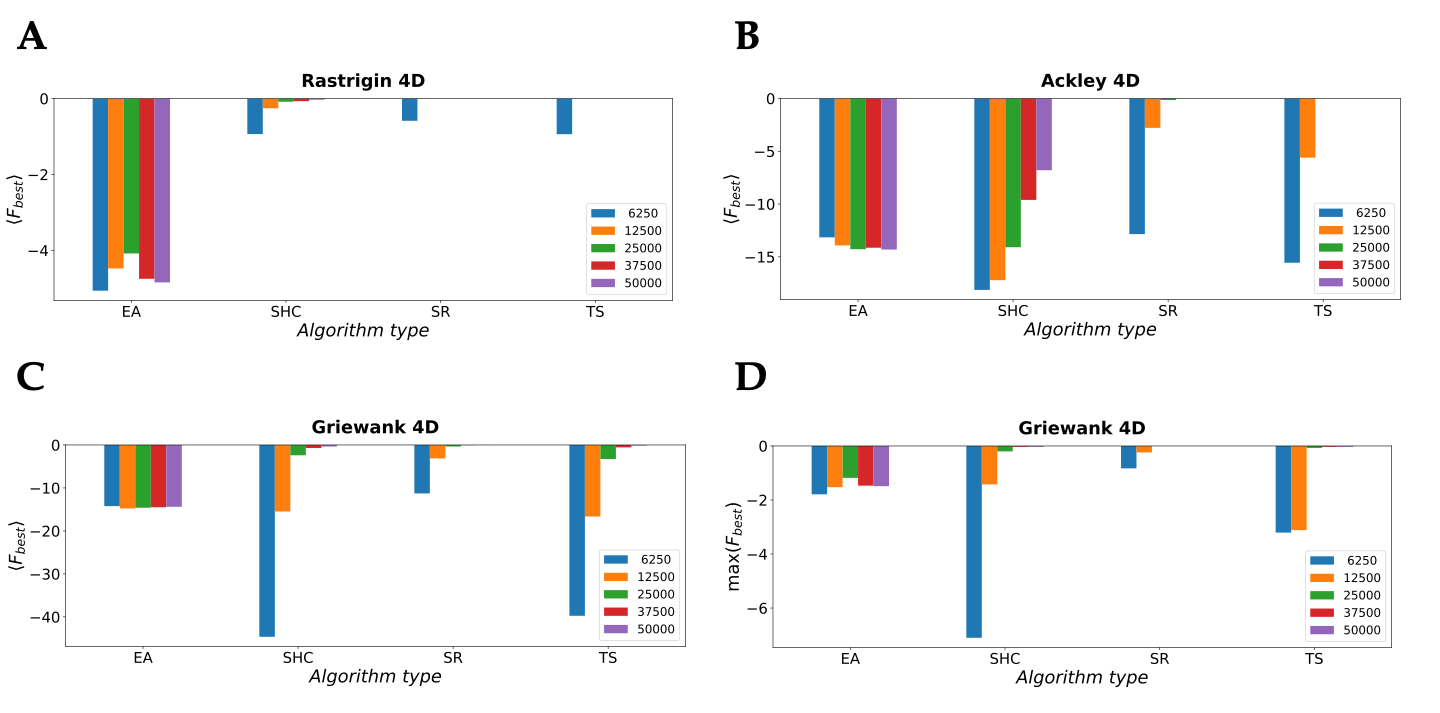}
\caption{
\textbf{Comparison of the algorithms conditioned on the number of unique fitness function calls: \textit{nnb} moveset.}
The maximum number of allowed function calls was set to $\{ 6250, 12500, 25000, 37500, 50000 \}$ for all algorithms.
In panels A-C we show the best fitness values found in each run averaged over $100$ independent runs with randomly chosen starting states.
(A) Rastrigin function.
(B) Ackley function.
(C) Griewank function.
(D) Same as C but for the globally best fitness values obtained over all $100$ runs instead of the averages.
EA -- Evolutionary Algorithm ($\mu = 0.1$, $r_x = 0.1$, $N_\text{pop} = 50$),
SHC -- Stochastic Hill Climbing ($T = 0.5$),
SR -- SmartRunner ($l_\mathrm{max} = 2$, $\xbar{R}^\mathrm{init} = 0.1$, $\alpha = 1.0$ for Rastrigin and Ackley, $\alpha = 10.0$ for Griewank),
TS -- Taboo Search ($L_\text{tabu} = 500$).
}
\label{fig:comparo_runs:nnb}
\end{figure}

\vspace{0.4cm}
\noindent
\textbf{Comparison of global optimization algorithms.} 
Different global optimization algorithms use different notions of a single step. While SA, SHC and SmartRunner define a single stochastic trial as an elementary step, a TS step involves querying all nearest neighbors, and an EA step involves rebuilding a population subjected to crossover and mutation. To ensure a fair comparison, we have allocated a fixed number of \emph{novel} fitness function evaluations to each algorithm
and observed the resulting performance (SA had to be left out because its performance depends on the cooling schedule, such that stopping SA at $T > T_f$ puts it at an unfair disadvantage).
We note that with the \textit{nnb} moveset, SmartRunner consistently shows the best performance (Fig.~\ref{fig:comparo_runs:nnb}). As expected, the worst performance with this moveset is exhibited by EA as it is
unable to utilize more and more function evaluations to find states with better fitness -- in fact, EA often uses fewer function evaluations than was allocated to it, terminating instead when
the maximum number of steps is exceeded.

%With the \textit{spmut} moveset, SmartRunner's performance is not always top-ranked (Fig.~S16). Using the average of the best-fitness values as a measure of performance,
%SmartRunner is slightly behind TS for the Rastrigin function, outperforming EA and SHC by a large margin.

%%%%%%

\vspace{0.4cm}
\noindent
\textbf{SmartRunner tests on SK spin glass and Kauffman's NK models: quenched disorder.}
Next, we have turned to two challenging discrete-state systems with complex fitness landscapes. One is the Sherrington-Kirkpatrick (SK) spin glass model~\cite{Sherrington:1975}, with $N$ $\pm 1$ spins
coupled by random interactions that are independently sampled from the standard Gaussian distribution (SI Methods). The other is Kauffman's NK model used in evolutionary theory~\cite{Kauffman1989, Kauffman1993}, in which each of the $N$ $(0,1)$ sites interacts with $0 \le K \le N-1$ other sites chosen by random sampling. The fitness function for a given binary sequence is a sum over $N$ single-site contributions;
each single-site contribution is obtained by sampling from the standard uniform distribution (SI Methods). The model parameter $K$ serves to tune the degree of landscape ruggedness: the number of local
maxima increases rapidly as $K$ goes up. In both systems, the moveset consists of changing the binary state at a single site. Thus, each of the $2^N$ states has $N$ nearest neighbors. 
First, we focus on systems with quenched disorder, where random parameters of the system are generated only once and subsequently used in all
comparisons of global optimization algorithms.

We have carried out a SmartRunner hyperparameter search for the SK model with $N=200$ spins (Fig.~S16). We find that among all of the values tried, $\alpha = 0.1$ is clearly preferable (Fig.~S16A,C), with
a statistically significant improvement in performance (Fig.~S16E). On the other hand, the dependence on $\xbar{R}_\mathrm{~\!\!init}$ is very weak. As expected, the number of function evaluations
increases with $\alpha$ as novel states are explored more frequently (Fig.~S16B,D). The same conclusions are reached with the NK model with $N=200$ sites and $K=8$ couplings per site, with
$\alpha = 0.1$ identified again as the optimal value (Fig.~S17). We have also explored the $\alpha$ settings in models with 200, 500, and 1000 spins/sites (Fig.~S18). While $\alpha = 0.1$ is confirmed as
the optimal choice for $N = 200$ models, $\alpha = 0.01$ is preferable for $N = 500,1000$.

%%%%
\begin{table}[!htb]
%\begin{tabular}{ |p{8cm}|p{5cm}|  }
\small
\begin{tabular}{ |l|l|l|l|l|l|l| }
 \multicolumn{6}{c}{$\langle \mathcal{F}_{best} \rangle \pm \sigma_{\mathcal{F}_{best}}$} \\
 \hline
 %\hline
 $\mathbf{N_s}$ & $\mathbf{SR}$ & $\mathbf{TS}$ & $\mathbf{SA}$  &  $\mathbf{SHC}$ & $\mathbf{EA}$ \\
 \hline
 $\mathbf{200}$ &  $0.715 \pm 0.012$ &  $0.711 \pm 0.009$ & $\mathbf{0.723 \pm 0.003}$ & $0.719 \pm 0.009$ & $0.688 \pm 0.020$ \\
 \hline
 $\mathbf{500}$ &  $\mathbf{0.746 \pm 0.008}$ & $0.722 \pm 0.009$ &  $0.710 \pm  0.010$ & $0.735 \pm 0.008$ & $0.668 \pm 0.010$ \\
\hline
$\mathbf{1000}$ & $\mathbf{0.733  \pm 0.005}$ & $0.684 \pm 0.010$ & $0.492 \pm 0.015$ & $0.557 \pm 0.006$ & $0.598 \pm 0.009$ \\
\hline
\multicolumn{6}{c}{} \\
\multicolumn{6}{c}{$\text{max}(\mathcal{F}_{best})$} \\
\hline
$\mathbf{N_s}$ & $\mathbf{SR}$ & $\mathbf{TS}$ & $\mathbf{SA}$  &  $\mathbf{SHC}$ & $\mathbf{EA}$ \\
 \hline
$\mathbf{200}$ & $\mathbf{0.727}$  & 0.725  & $\mathbf{0.727}$  &  $\mathbf{0.727}$ &  0.717 \\
\hline
$\mathbf{500}$ & $\mathbf{0.757}$ &  0.734 &  0.729 &   0.744 &  0.684 \\
\hline
$\mathbf{1000}$ & $\mathbf{0.740}$ &  0.704 &  0.517 &  0.566 &  0.614 \\
\hline
\end{tabular}
\caption{
\textbf{Comparison of the algorithm performance on the SK model.} $\langle \mathcal{F}_{best} \rangle$ is the average of the best-fitness values found in each run, averaged over $10$ independent runs with randomly chosen starting states;  $\sigma_{\mathcal{F}_{best}}$ is the corresponding standard deviation;  $\text{max}(\mathcal{F}_{best})$ is the largest of the best-fitness values;
N$_\text{s}$ is the number of spins.
SR -- SmartRunner ($l_\mathrm{max} = 2$, $\xbar{R}^\mathrm{init} = 0.01$, $\alpha = 0.01$), TS -- Taboo Search ($L_\text{tabu} = 5000$),
SA  -- Simulated Annealing ($T_i = 0.01$, $T_f = 0.001$, linear cooling schedule), SHC -- Stochastic Hill Climbing ($T = 10^{-3}$),
EA -- Evolutionary Algorithm ($\mu = 0.2$, $r_x = 0.5$, $N_\text{pop} = 100$). In SR, SA and SHC the total number of steps $l_\mathrm{tot} = 1.5 \times 10^6, 10^6, 5 \times 10^5$
for the models with 200, 500 and 1000 spins, respectively. In TS, the total number of steps is rescaled by the number of nearest neighbors ($l_\mathrm{tot} = 7.5 \times 10^3, 2 \times 10^3, 5 \times 10^2$);
in EA, the total number of steps is rescaled by the population size ($l_\mathrm{tot} = 1.5 \times 10^4, 10^4, 5 \times 10^3$).
The best result in each row is highlighted in boldface. For consistency, all runs employed a single random realization of the SK model (quenched disorder).
}
\label{table:SK}
\end{table}

%%%%
\begin{table}[!htb]
%\begin{tabular}{ |p{8cm}|p{5cm}|  }
\small
\begin{tabular}{ |l|l|l|l|l|l|l| }
 \multicolumn{6}{c}{$\langle \mathcal{F}_{best} \rangle \pm \sigma_{\mathcal{F}_{best}}$} \\
 \hline
 %\hline
 $\mathbf{N_s}$ & $\mathbf{SR}$ & $\mathbf{TS}$ & $\mathbf{SA}$  & $\mathbf{SHC}$ & $\mathbf{EA}$ \\
 \hline
 $\mathbf{200}$ &  $0.778 \pm 0.005$ &  $0.772 \pm 0.005$ & $0.771 \pm 0.005$ & $\mathbf{0.785 \pm 0.005}$ & $0.751 \pm 0.008$ \\
 \hline
 $\mathbf{500}$ &  $\mathbf{0.776 \pm 0.003}$ & $0.748 \pm 0.007$ &  $0.675 \pm 0.005$ & $0.702 \pm 0.003$ & $0.727 \pm 0.007$ \\
\hline
$\mathbf{1000}$ & $\mathbf{0.770  \pm 0.001}$ & $0.732 \pm 0.004$ & $0.601 \pm 0.004$ & $0.616 \pm 0.002$ & $0.700 \pm 0.005$ \\
\hline
\multicolumn{6}{c}{} \\
\multicolumn{6}{c}{$\text{max}(\mathcal{F}_{best})$} \\
\hline
$\mathbf{N_s}$ & $\mathbf{SR}$ & $\mathbf{TS}$ & $\mathbf{SA}$  &  $\mathbf{SHC}$ & $\mathbf{EA}$ \\
 \hline
$\mathbf{200}$ & 0.789  & 0.781  & 0.782  & $\mathbf{0.795}$ &  0.761 \\
\hline
$\mathbf{500}$ & $\mathbf{0.784}$ &  0.758 &  0.683 & 0.706 & 0.739 \\
\hline
$\mathbf{1000}$ & $\mathbf{0.772}$ &  0.738 &  0.606 & 0.620 & 0.708 \\
\hline
\end{tabular}
\caption{
\textbf{Comparison of the algorithm performance on the NK model.} N$_\text{s}$ is the number of sites (each site has $8$ randomly chosen intra-sequence couplings per site);
all other quantities and parameter settings are as in Table~\ref{table:SK}. For consistency, all runs employed a single random realization of the NK model (quenched disorder).
}
\label{table:NK}
\end{table}

Finally, we carry out a side-by-side comparison of the performance of all $5$ algorithms: SR, TS, SA, SHC, and EA on the SK models (Table~\ref{table:SK}) and the NK
models (Table~\ref{table:NK}). To mimic a realistic situation in which computer resources are a limiting factor and the fitness landscapes are exceedingly large,
we have chosen a single set of hyperparameter settings for each algorithm. Thus, SR was run with $\alpha = 0.01$, even though the above analysis shows that $\alpha = 0.1$ is in fact
a better choice for $N = 200$. The only exception to this rule is SHC, where we carried out a mini-scan over the values $T$ to optimize performance. All algorithms except for SmartRunner
were run on the original landscapes without occupancy penalties. For SA, $T_f \simeq 0$ should be reasonable, while $T_i = 1.0$ is dictated by the overall scale of the landscape.
For EA, a 3D scan over $\mu$, $r_x$, $N_\text{pop}$ is not feasible, so that we had to settle for `typical' values. Thus, more complex algorithms with several hyperparameters are implicitly
penalized, as they are likely to be in a realistic research setting.

We find that SmartRunner ranks the highest overall in this competition. For the SK models, it is in the second place for $N = 200$ and the first place for $N = 500,1000$ if judged by the
average of all solutions (Table~\ref{table:SK}). If judged by the globally best solution, the SmartRunner shares the first place with SHC for $N = 200$ and again takes
the first place for $N = 500,1000$. Similar results are seen with the NK model (Table~\ref{table:NK}): by both the average and the globally best measures, SmartRunner is second for
the $N=200$ model and first for the larger models with $500$ and $1000$ sites. The somewhat weaker performance of SmartRunner on the $N = 200$ systems could be improved by switching
to $\alpha = 0.1$ (Figs.~S16,S17). However, this would give SmartRunner an unfair advantage in the context of this competition, in which every algorithm was run with a single reasonable set
of hyperparameters.

\begin{figure}[!htb]
\centering
%\hspace*{-2.5cm}
\includegraphics[scale=0.12]{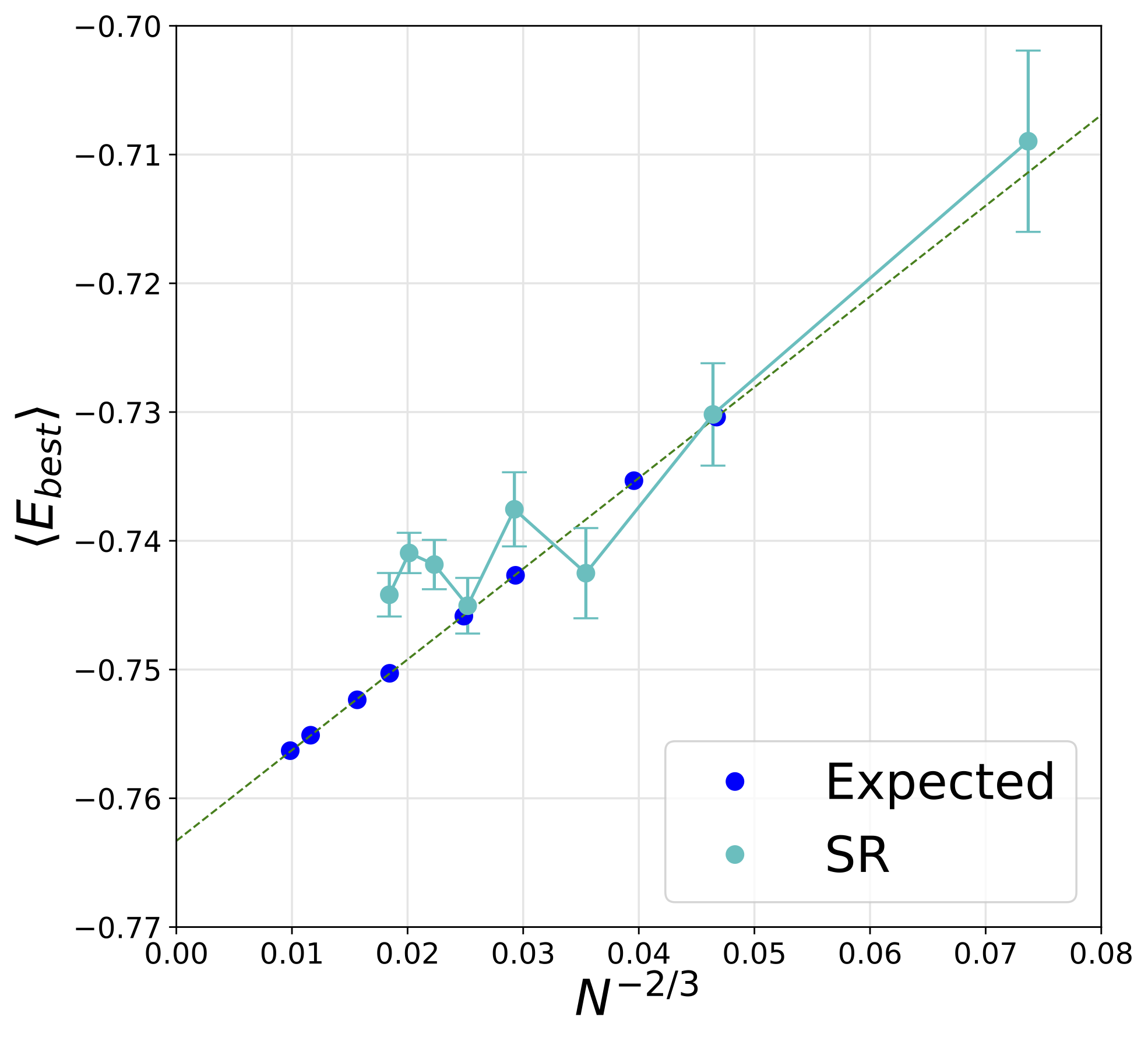}
\caption{
\textbf{Finite-size corrections to the ground state energy per spin in the SK model.}
Cyan dots: average of the best-energy values found by SmartRunner on SK landscapes with $N = 50, 100, 150, 200, 250, 300, 350, 400$ spins.
For each value of $N$, 18 to 23 independent runs with $l_\mathrm{max} = 2$, $\xbar{R}^\mathrm{init} = 0.01$, $\alpha = 0.1$
were carried out, each one with randomly generated spin couplings and starting from a random spin
configuration. The total number of steps $l_\mathrm{tot}$ ranged from $4.5 \times 10^6$ to $3.0 \times 10^7$ depending on $N$.
Error bars represent the errors of the mean.
Blue dots: numerical results for the finite-size corrections to the SK ground state energy reported by S. Boettcher using
the Extremal Optimization algorithm (Table 1 in Ref.~\cite{Boettcher:2005}).
Dashed green line: a linear fit to Boettcher's ground state energies yielding $\langle E_{best} (N) \rangle = \langle E_{best} (\infty) \rangle + m N^{-2/3}$, where
$m = 0.7047$ is the slope and $\langle E_{best} (\infty) \rangle = -0.7633$ is the asymptotic Parisi energy for the infinite system~\cite{Parisi:1980}.
All energies are divided by the number of spins $N$ to produce intensive quantities.
}
\label{fig:SKcomparo}
\end{figure}

\vspace{0.4cm}
\noindent
\textbf{Prediction of the SK ground state energies averaged over disorder.}
Next, we have investigated the ability of SmartRunner to reproduce finite-size corrections to average ground state energies of the SK model (Fig.~\ref{fig:SKcomparo}).
The ground state energy per spin averaged over random spin couplings is known theoretically to be $-0.7633$ in the $N \to \infty$ limit of the SK model~\cite{Parisi:1980},
with the $2/3$ scaling exponent for finite-size corrections (i.e., $\langle E_{best} (N) \rangle \sim N^{-2/3}$) available from both theoretical~\cite{Parisi:1993}
and numerical~\cite{Boettcher:2005} investigations. This provides a baseline against which SmartRunner's ability to find the global minima of the SK energy can be judged.
We find that the average ground-state energy per spin predicted by SmartRunner is reasonably close to the expected straight line in Fig.~\ref{fig:SKcomparo}, although there are
statistically significant deviations for the three largest systems ($N = 300, 350, 400$), indicating that SmartRunner does not quite reach the true ground states in these cases.
Overall, SmartRunner's performance on these systems is less reliable than that of Extremal Optimization, a heuristic algorithm specifically adapted to the SK model and requiring
a simplified  probabilistic model for spin couplings~\cite{Boettcher:2005,Boettcher:2010}.

\section*{Discussion and Conclusion}

In this work, we have developed a novel approach to global optimization called SmartRunner. Instead of relying on qualitative similarities with physical, chemical or biological systems,
SmartRunner employs an explicit probabilistic model for accepting or rejecting a move on the basis of the immediate previous history of the optimization process. The key quantity guiding
SmartRunner decisions is $p_f$, the probability of finding a higher-fitness target in the next random trial. This probability has nearly universal asymptotics
and can be effectively represented by a function that depends only on $n$, the number of previously rejected attempts to change the current state of the system. In other words, the dependence
of SmartRunner's behavior on such details of the systems as the number of nearest neighbors and the transition rates is fairly weak, making our approach
applicable to a wide range of objective functions and movesets. Overall, SmartRunner can be viewed as an adaptive search policy designed to maximize fitness gain per step.

Interestingly, SmartRunner's global optimization policy amounts to hill ascent on a fitness landscape modified with an easily computed adaptive occupancy penalty.
The occupancy penalty makes rejecting moves less and less favorable as the number of unsuccessful attempts to change the current state grows.
Ultimately, one of the nearest neighbors is accepted even if the step is deleterious on the original fitness landscape. This behavior allows SmartRunner to climb out of local basins of attraction
(Fig.~\ref{fig:2Dsearch}). In principle, the adaptive fitness landscape given by Eq.~\eqref{Ftilde} can be explored using any global optimization algorithm.

We have tested SmartRunner's performance on a standard set of functions routinely used to evaluate the performance of global optimization algorithms~\cite{Torn:1989}.
These 4D functions are characterized by numerous local maxima that make it challenging to find a single global maximum.
We find that SmartRunner exhibits the highest fitness gain per novel fitness function evaluation compared to three other state-of-the-art gradient-free algorithms (Fig.~\ref{fig:comparo_runs:nnb}).
This is especially important in situations where fitness function calls are computationally expensive.
Interestingly, when adaptive fitness landscapes were given as input to other global optimization algorithms, the best results were obtained when the other algorithms' policy for accepting and rejecting
moves closely resembled the SmartRunner policy of hill climbing on the modified fitness landscape (Fig.~\ref{fig:tunnel}).
For example, with simulated annealing the globally best strategy was to set the initial temperature to a very low value, essentially reducing simulated annealing to hill ascent.
Finally, we observe that the SmartRunner approach is flexible enough to adapt to substantial changes in the moveset, from $\mathcal{O} (10^0)$ local moves to $\mathcal{O} (10^{2} - 10^{3})$
random updates of a single randomly chosen coordinate (Figs.~\ref{fig:SRruns:R},\ref{fig:SRruns:Rspmut}).

We have also tested SmartRunner on two challenging models with long-range couplings and multiple local minima or maxima: the Sherrington-Kirkpatrick spin glass model~\cite{Sherrington:1975} and the
Kauffman's NK model of fitness~\cite{Kauffman1989, Kauffman1993}. In systems with quenched disorder, SmartRunner performs very well compared with four other general-purpose
global optimization algorithms (Tables~\ref{table:SK},\ref{table:NK}). It is also fairly reliable in locating ground-state energies averaged over disorder in the SK model, although the results are
inferior to those obtained by Extremal Optimization, a heuristic algorithm specifically adapted to finding the ground states in the SK model~\cite{Boettcher:2005,Boettcher:2010} (Fig.~\ref{fig:SKcomparo}).

In summary, SmartRunner implements a novel global optimization paradigm which offers a viable alternative to current algorithms.
The SmartRunner approach described here works on discrete or discretized fitness landscapes and does not make use of the gradient of the objective function
in implementing its stochastic policy. In the future, we intend to adapt SmartRunner to carry out global optimization on continuous landscapes where the gradient of the objective
function can be computed efficiently. Such optimization will be of great interest in modern machine learning. For example, training artificial neural networks relies on the differentiability
of objective functions and optimization methods based on stochastic gradient descent~\cite{Goodfellow2016,Mehta2019}, which may get trapped in local minima.

\section*{Software Availability}

The Python3 code implementing SmartRunner and four other gradient-free global optimization algorithms discussed here is available at
\texttt{https://github.com/morozov22/SmartRunner}.

\section*{Acknowledgements}

We gratefully acknowledge illuminating discussions with Stefan Boettcher.
JY and AVM were supported by a grant from the National Science Foundation (NSF MCB1920914).

\clearpage

\bibliographystyle{nar}
\bibliography{sample,optimization}

\end{document}

\maketitle

\newpage
\section*{Supplementary Methods}

\subsection*{Standard test functions}

We employ 3 standard test functions~\cite{Torn:1989}, including the 4D Rastrigin function:
\begin{equation} \label{eq:Rdef}
\mathcal{F} (\vec{\bf x}) = -4 - \sum_{i=1}^{4} \big(x_i^2 - \cos (18 x_i) \big),~x_i \in [-5,5], \forall i,
\end{equation}
the 4D Ackley function:
\begin{equation} \label{eq:Adef}
\mathcal{F} (\vec{\bf x}) = -20 - e + 20 \exp \Big( -0.2 \sqrt{\frac{1}{4} \sum_{i=1}^{4} x_i^2} \Big) +
\exp \Big( \frac{1}{4} \sum_{i=1}^{4} \cos (2 \pi x_i) \Big),~x_i \in [-32.8,32.8], \forall i,
\end{equation}
and the 4D Griewank function:
\begin{equation} \label{eq:Gdef}
\mathcal{F} (\vec{\bf x}) = -1 - \frac{1}{4000} \sum_{i=1}^{4} x_i^2 + \prod_{i=1}^{4} \cos \big(\frac{x_i}{\sqrt{i}}\big),~x_i \in [-600,600], \forall i
\end{equation}

All three functions have multiple local maxima and a unique global maximum located at $\vec{\bf x} = \vec{\bf 0} $ ($\mathcal{F} (\vec{\bf 0}) = 0$).
The fitness landscapes are discretized, with periodic boundary conditions and $\Delta x = (0.05, 0.2, 1.0)$ steps in all four directions for Rastrigin, Ackley and Griewank functions,
respectively.

\subsection*{Sherrington-Kirkpatrick spin glass model}

Consider a system of $N$ spins which can point either up or down: $s_i = \pm 1$, $i = 1 \dots N$.
The Sherrington-Kirkpatrick (SK) model is defined by the Hamiltonian~\cite{Sherrington:1975}:
\begin{equation} \label{eq:H_SK}
\mathcal{H}_\text{SK} (s) = - \frac{1}{\sqrt{N}} \sum_{1 \le i < j \le N} J_{ij} s_i s_j,
\end{equation}
where $s = (s_1, s_2, \dots, s_N)$ and $J_{ij}$ are random spin couplings independently sampled from a standard Gaussian distribution:
\begin{equation} \label{eq:Jij}
P(J_{ij}) = \frac{1}{\sqrt{2 \pi}} e^{-\frac{J_{ij}}{2}}.
\end{equation}

The SK model is characterized by a large number of local minima and therefore presents a challenging problem to global optimization algorithms.
The ground state energy per spin of the SK model (the global minimum) asymptotically approaches $\mathcal{H}_\text{SK}/N \to -0.7633$ in the $N \to \infty$ limit~\cite{Parisi:1980}.
%However, this result requires integrating over the random spin couplings in the thermodynamic limit and therefore can only serve as a rough estimate for finite systems
%with quenched disorder that we study here (in systems with quenched disorder, random couplings are generated only once and subsequently used in all
%comparisons of global optimization algorithms).
All SK energies are divided by the number of spins $N$ to produce intensive quantities.

\subsection*{Kauffman's NK model}

In Kauffman's NK model~\cite{Kauffman1989, Kauffman1993}, each of the $N$ sites in the gene (or genes in the genome) interacts with $K$ other sites chosen by random sampling.
The fitness of the genotype $s = (s_1, s_2, \dots, s_N)$ ($s_i = \{0,1\}$, $i = 1 \dots N$) is given by
\begin{equation} \label{eq:F_NK}
\mathcal{F}_\text{NK} (s) = \sum_{\mu=1}^N \mathcal{F}_\mu(s_\mu, s_{n_1(\mu)}, \dots, s_{n_K(\mu)}),
\end{equation}
where $n_1(\mu), \dots, n_K(\mu)$ are $K$ randomly-chosen interaction partners of the site $s_\mu$. The single-site fitnesses $\mathcal{F}_\mu$ are obtained by sampling from a standard uniform distribution
($U(0,1)$); each combination of the $2^{K+1}$ possible states of the argument corresponds to an independent sample from this distribution.
When $K=0$, the NK landscape becomes fully additive. Because in this limit the landscape is smooth and has a single maximum, it is sometimes called the ``Mount Fuji'' model~\cite{Aita2000}. The amount of landscape ruggedness can be tuned by increasing $K$ to the maximum value of $N-1$. With $K = N-1$, the fitnesses of different sequences are uncorrelated; this model is called the ``House of Cards''~\cite{Kingman1978} due to the unpredictable fitness effects of mutations. Numerous results are available for the statistical mechanics of NK
landscapes~\cite{Kauffman1989,Flyvbjerg1992,Altenberg1997,Rokyta2006,Kryazhimskiy2009}. 
In particular, in the $K = N-1$ limit the average number of local maxima is $2^L/(L + 1)$ for the binary alphabet considered here~\cite{Szendro2013}.
As with the SK model, we use NK models with quenched disorder -- for given values of $N$ and $K$,
all interaction partners and single-site fitnesses are generated once and then used in all subsequent comparisons of global optimization algorithms.
All fitnesses are divided by the number of sites $N$ to produce intensive quantities.

\subsection*{Alternative global optimization algorithms}

We have implemented four alternative global optimization algorithms: Simulated Annealing (SA), Stochastic Hill Climbing (SHC), Evolutionary Algorithm (EA), and
Taboo Search (TS) using Solid, a gradient-free Python optimization package (\texttt{https://github.com/100/Solid}). The modified Solid code for these four algorithms, implementing
fitness functions augmented with the occupancy penalties, is available as part of the SmartRunner GitHub distribution package (\texttt{https://github.com/morozov22/SmartRunner}).
SA, SHC and TS search strategies were left exactly as implemented in Solid, while in EA fitnesses of all population members were made positive by subtracting the minimum fitness in the current population;
these fitnesses were then used to compute selection probabilities. Apart from this change, Solid's EA crossover and mutation strategies remained the same as in the original package.

%\begin{itemize}

%\item

%\end{itemize}

\newpage
\section*{Supplementary Figures}

%%%%%%
\begin{figure}[!htb]
\centering
%\hspace*{-2.5cm}
\includegraphics[scale=0.55]{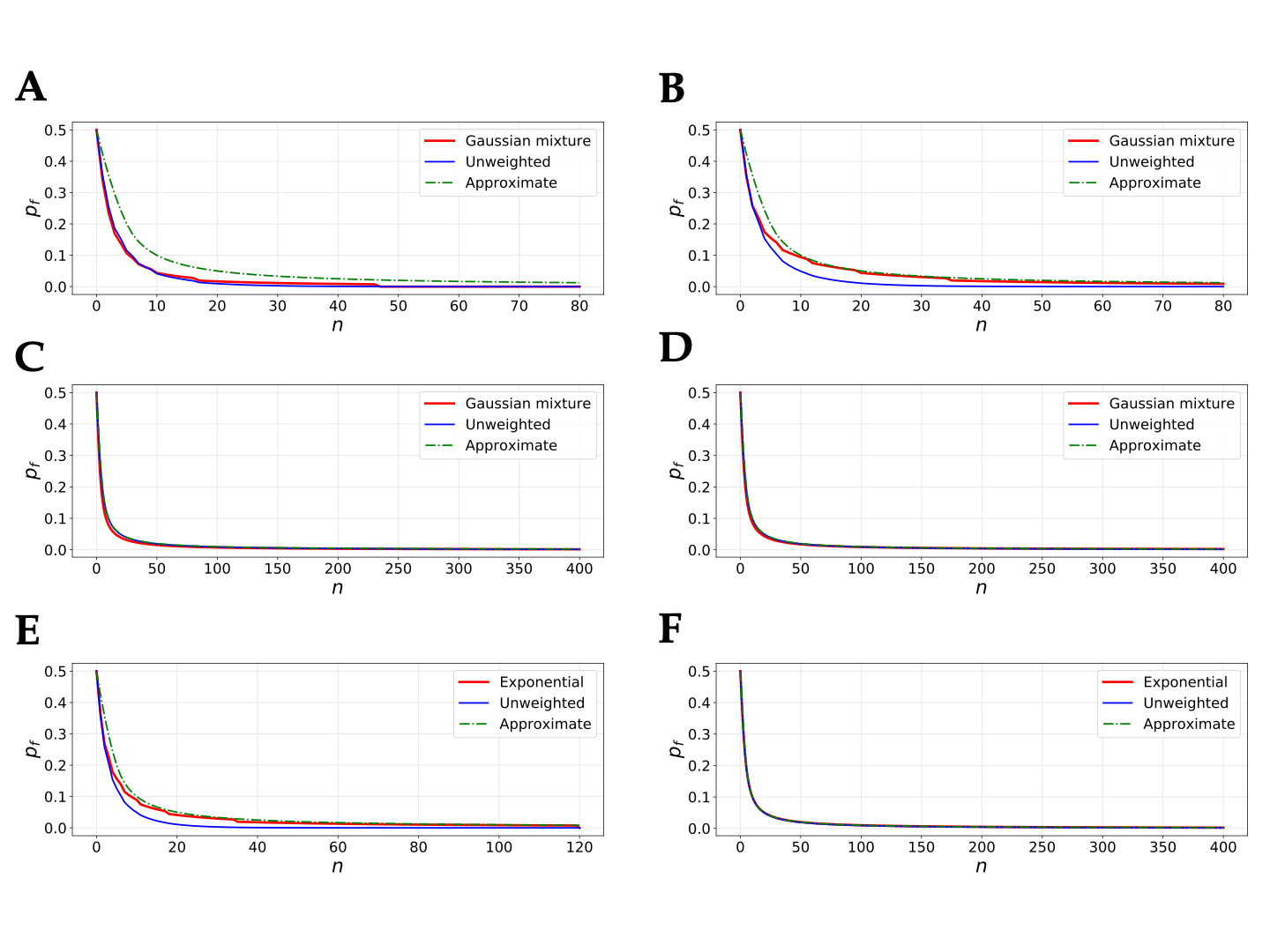}
\caption{
\textbf{The probability $p_f$ of finding an unobserved higher-fitness target as a function of the number of trials $n$.}
In all panels, solid blue lines represent $p_f$ in systems with equal edge weights (Eq.~(10) in the main text), while dashed-dotted green lines show the approximate $p_f$
expression (Eq.~(23) in the main text). In panels A-D, solid red lines represent systems with edge weights distributed according to a two-component Gaussian mixture model ($\mathcal{P} = 2$
in Eq.~(17) in the main text); $p_f$ is computed by substituting $\beta$ from Eq.~(21) instead of $\gamma$ in Eq.~(10). In panels E-F, solid red lines correspond to systems with
exponentially distributed edge weights (Eq.~(22) in the main text), with $p_f$ computed using $\beta = \log \left( 1 + \frac{n}{\mathcal{N}} \right)$
($\mathcal{N}$ is the total number of nearest neighbors) instead of $\gamma$ in Eq.~(10). 
Note that computing $p_f$ in systems with weighted or unweighted edges requires the knowledge of $m_p$, the number of nearest neighbors of the current node visited once or more by the random walker.
In contrast, computing $p_f$ with the approximate expression (Eq.~(23)) imposes no such requirement. Since $m_p$ is a stochastic variable, we use its expected value $E[m_p]$ in the plots,
computed using $\beta$ from Eq.~(21) instead of $\gamma$ in Eq.~(7) in the main text for the Gaussian mixture distribution, and $E[m_p] = \frac{\mathcal{N} n}{\mathcal{N} + n}$ for the 
exponential distribution. To enable direct comparisons, the same $E[m_p]$ values are used in both weighted and unweighted $p_f$ computations in each panel.
(A) Gaussian mixture model with $(p_1, p_2) = (0.5, 0.5)$, $(\bar{w}_1, \bar{w}_2) = (0.1, 0.5)$, $(\sigma_1, \sigma_2) = (0.2, 0.2)$; $\mathcal{N} = 8$.
(B) Gaussian mixture model with $(p_1, p_2) = (0.8, 0.2)$, $(\bar{w}_1, \bar{w}_2) = (0.1, 1.0)$, $(\sigma_1, \sigma_2) = (0.1, 0.1)$; $\mathcal{N} = 8$.
(C) Same as (A) but with $\mathcal{N} = 500$.
(D) Same as (B) but with $\mathcal{N} = 500$.
(E) Exponential model with $\mathcal{N} = 8$.
(F) Exponential model with $\mathcal{N} = 500$.
}
\label{fig:6curves}
\end{figure}

%%%%%%
%\begin{figure}[!htb]
%\centering
%%\hspace*{-2.5cm}
%\includegraphics[scale=0.55]{Figures/Figure_2Dpics.png}
%\caption{
%\textbf{SmartRunner search for the global maximum on a 2D fitness landscape with two basins of attraction.}
%The fitness function is given by a weighted sum of two Gaussians:
%$\mathcal{F} (x,y) = \sum_{k=1}^{2} A_k \exp \left[ -{(x - \bar{x}_k)^2}/{2 \sigma_{x,k}^2} - {(y - \bar{y}_k)^2}/{2 \sigma_{y,k}^2} \right]$, with
%$(A_1, A_2) = (50, 75)$, $(\bar{x}_1, \bar{y}_1) = (-3.5, 0.0)$, $(\sigma_{x,1}, \sigma_{y,1}) = (3.0, 2.0)$,
%$(\bar{x}_2, \bar{y}_2) = (3.5, 0.0)$, and $(\sigma_{x,2}, \sigma_{y,2}) = (2.0, 3.0)$.
%The fitness landscape is discretized with a $0.01$ step in both $x$- and $y$-directions, with $x \in [-10,10]$, $y \in [-10,10]$ and
%periodic boundary conditions; the total number of states is $4 \times 10^6$.
%Panels A-D show $4$ SmartRunner trajectories starting from $(x_0, y_0) = (-8.0, 0.0)$, with $l_\mathrm{tot} = 10^4, 10^5, 5 \times 10^5, \text{ and } 10^6$,
%respectively. The trajectories are color-coded from dark blue to yellow as the run progresses. All runs used $\alpha = 0.1$, $l_\mathrm{max} = 2$,  and $\xbar{R}^\mathrm{init} = 0.1$.
%The best fitness value found in the run is $\mathcal{F}_\mathrm{best} = 50.17$ (left peak) in panels A-C and $78.48$ (right peak) in panel D.
%}
%\label{fig:2Dsearch}
%\end{figure}

%%%%%%
\begin{figure}[!htb]
\centering
%\hspace*{-2.5cm}
\includegraphics[scale=0.60]{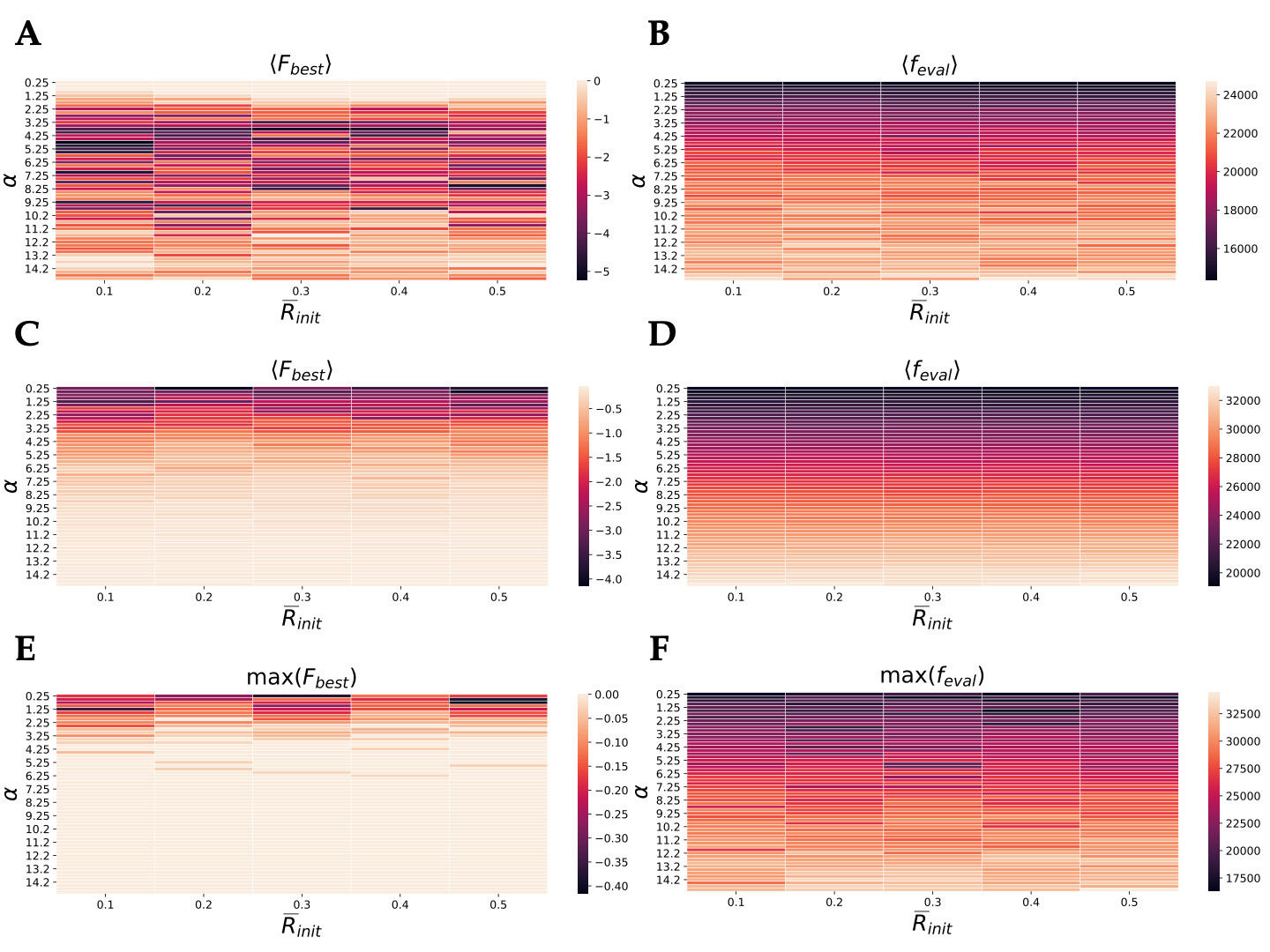}
\caption{
\textbf{SmartRunner exploration of the Ackley and Griewank test functions: \textit{nnb} moveset.}
(A) A scan over SmartRunner hyperparameters ($l_\mathrm{max} = 2$, $nnb$ moveset) for the Ackley function:
the initial value of the expected rate of fitness gain per step $\xbar{R}_\mathrm{~\!\!init}$ and the level of optimism $\alpha$.
Each cell in the heatmap represents best fitness values found in each run, averaged over $50$ independent runs with $l_\mathrm{tot} = 10^5$ steps each and randomly chosen starting states.
(B) Same as A but with the average taken over the number of fitness function evaluations (unique fitness function calls) in each run.
(C) Same as A but for the Griewank function.
(D) Same as B but for the Griewank function.
(E) Same as C but with the globally best fitness value obtained over all $50$ runs shown instead of the average.
(F) Same as D but with the number of fitness function evaluations corresponding to the globally best run from E shown instead of the average over $50$ runs.
}
\label{fig:SRruns:AG}
\end{figure}

%%%%%%
%\begin{figure}[!htb]
%\centering
%%\hspace*{-2.5cm}
%\includegraphics[scale=0.60]{Figures/Figure_Rastrigin4D_spmut.png}
%\caption{
%\textbf{SmartRunner exploration of the Rastrigin test function: \textit{spmut} moveset.}
%Same as Fig.~2 in the main text (including SmartRunner settings and hyperparameter value settings in panels C-E) but with the $spmut$ moveset.
%}
%\label{fig:SRruns:Rspmut}
%\end{figure}

%%%%%%
\begin{figure}[!htb]
\centering
%\hspace*{-2.5cm}
\includegraphics[scale=0.60]{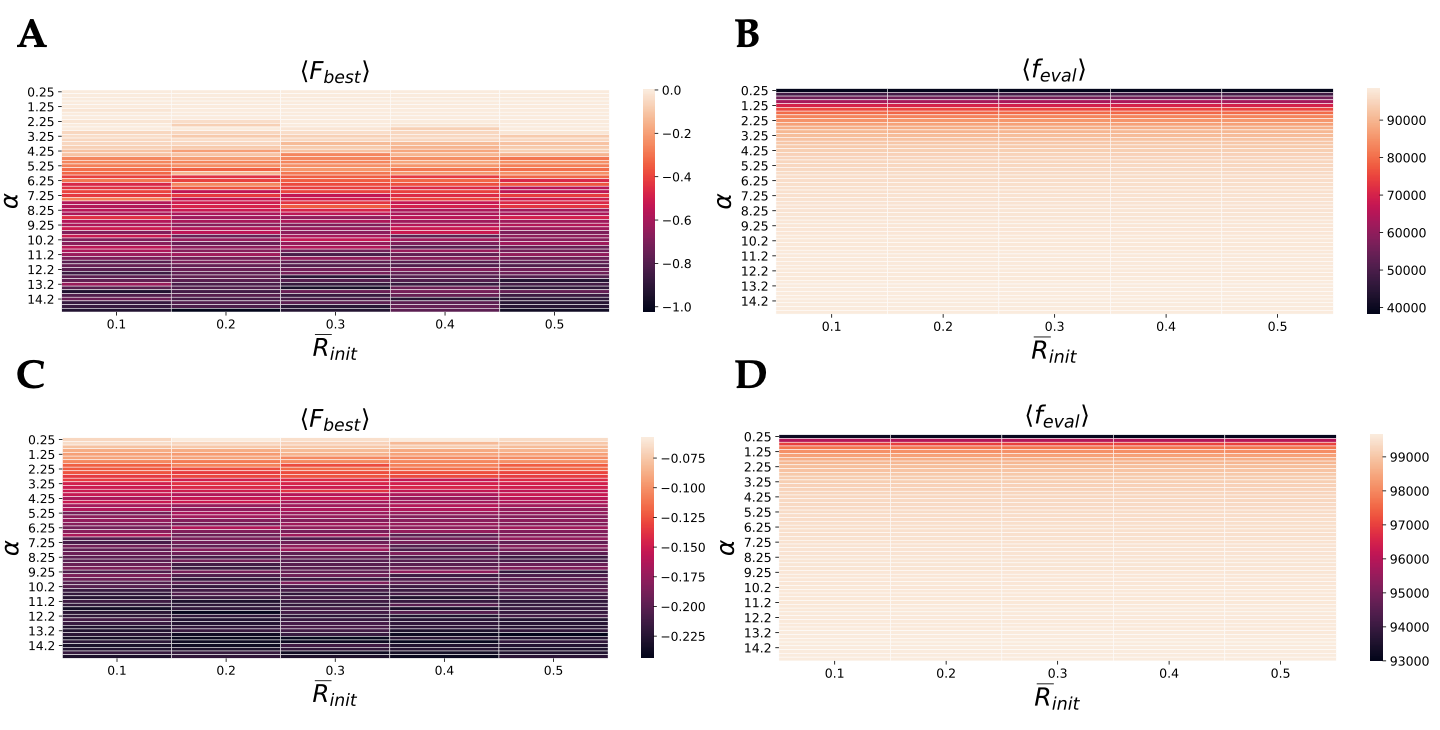}
\caption{
\textbf{SmartRunner exploration of the Ackley and Griewank test functions: \textit{spmut} moveset.}
Same as Fig.~\ref{fig:SRruns:AG}A-D (including SmartRunner settings) but with the $spmut$ moveset.
}
\label{fig:SRruns:AGspmut}
\end{figure}

%%%%%%
\begin{figure}[!htb]
\centering
%\hspace*{-2.5cm}
\includegraphics[scale=0.55]{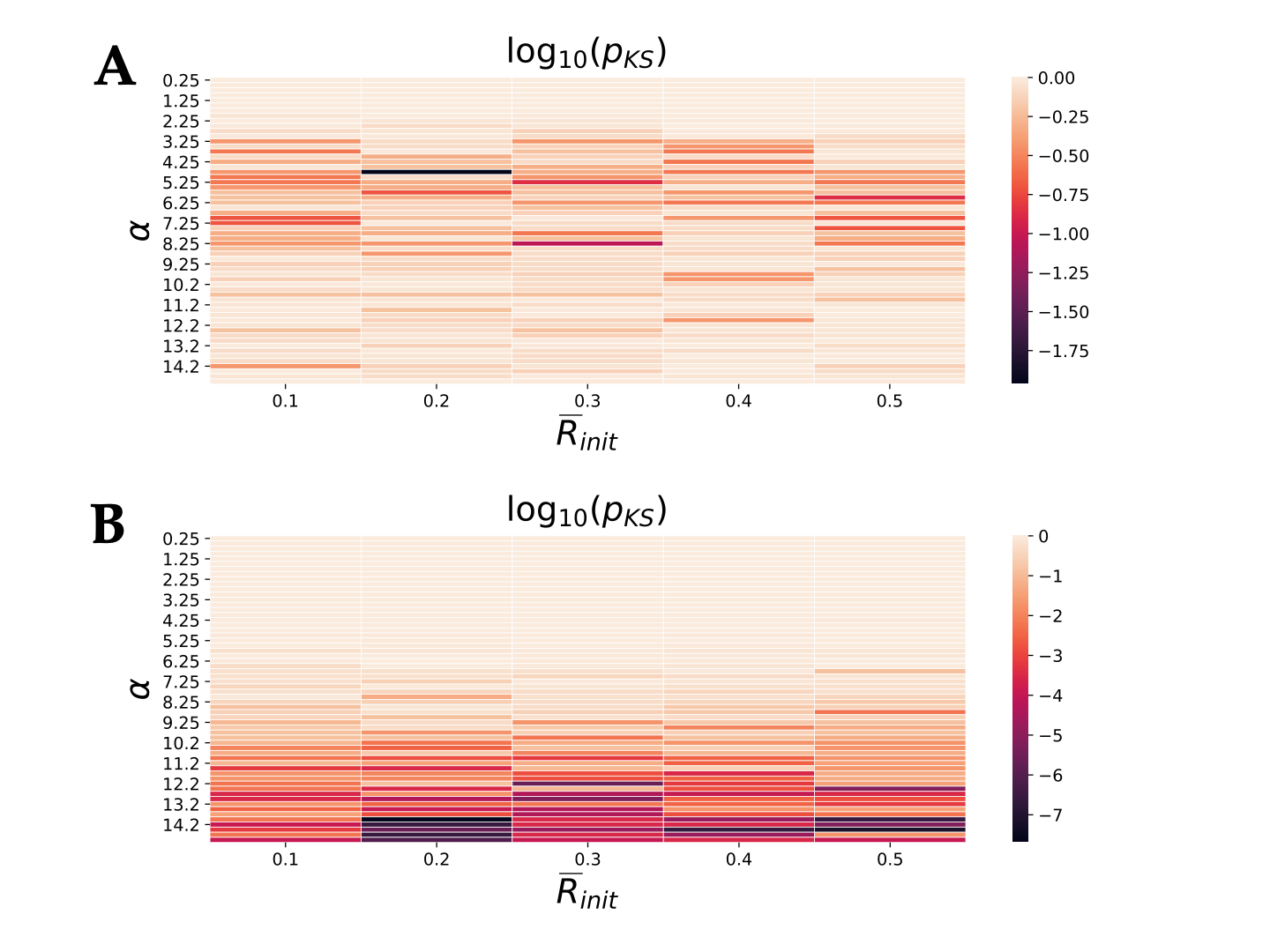}
\caption{
\textbf{Kolmogorov-Smirnov (KS) p-value analysis of the SmartRunner hyperparameter settings: Rastrigin test function.}
(A) One-sided KS tests were used to investigate the significance of the differences between two distributions of best-fitness values: the heatmap cell with the highest average of best-fitness values
(over $50$ independent SmartRunner runs) vs. every other cell. Low p-values ($< 0.05$) indicate that the cell with the highest average has a significantly better distribution of best-fitness values than the other cell.
High p-values ($\ge 0.05$) indicate that the cell with the highest average has a distribution of best-fitness values that is either indistinguishable from, or worse than the distribution in the other cell.
SmartRunner was used with the $nnb$ moveset.
(B) Same as A but with the $spmut$ moveset.
}
\label{fig:SRruns:pKSself}
\end{figure}

%%%%%%
\begin{figure}[!htb]
\centering
%\hspace*{-2.5cm}
\includegraphics[scale=0.60]{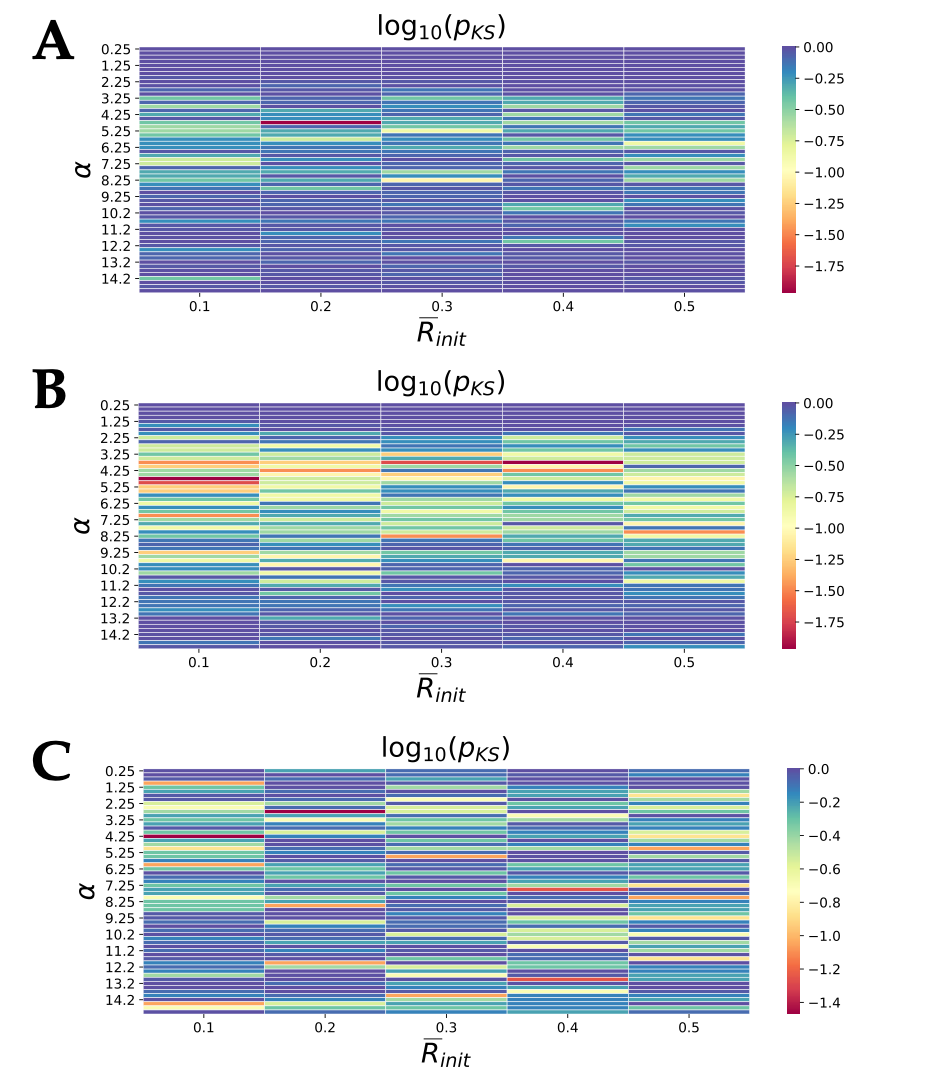}
\caption{
\textbf{Kolmogorov-Smirnov (KS) p-value analysis of the SmartRunner $l_\mathrm{max}$ hyperparameter setting.}
(A) Rastrigin test function. One-sided KS tests were used to investigate the significance of the differences between distributions of best-fitness values yielded by 
SmartRunner runs with $l_\mathrm{max} = 3$ and $l_\mathrm{max} = 2$. Each KS test compared two best-fitness distributions with the same values of $\alpha$ and
$\xbar{R}_\mathrm{~\!\!init}$.
Low p-values ($< 0.05$) indicate that $l_\mathrm{max} = 3$ runs have yielded a significantly better distribution of best-fitness values
than the runs with $l_\mathrm{max} = 2$.
High p-values ($\ge 0.05$) indicate that $l_\mathrm{max} = 3$ runs have yielded a distribution of best-fitness values that is either indistinguishable from,
or worse than the distribution with $l_\mathrm{max} = 2$.
(B) Same as A but for the Ackley test function.
(C) Same as A but for the Griewank test function.
All runs employed the $nnb$ moveset.
}
\label{fig:SRruns:lmax3_2}
\end{figure}

%%%%%%
\begin{figure}[!htb]
\centering
%\hspace*{-2.5cm}
\includegraphics[scale=0.55]{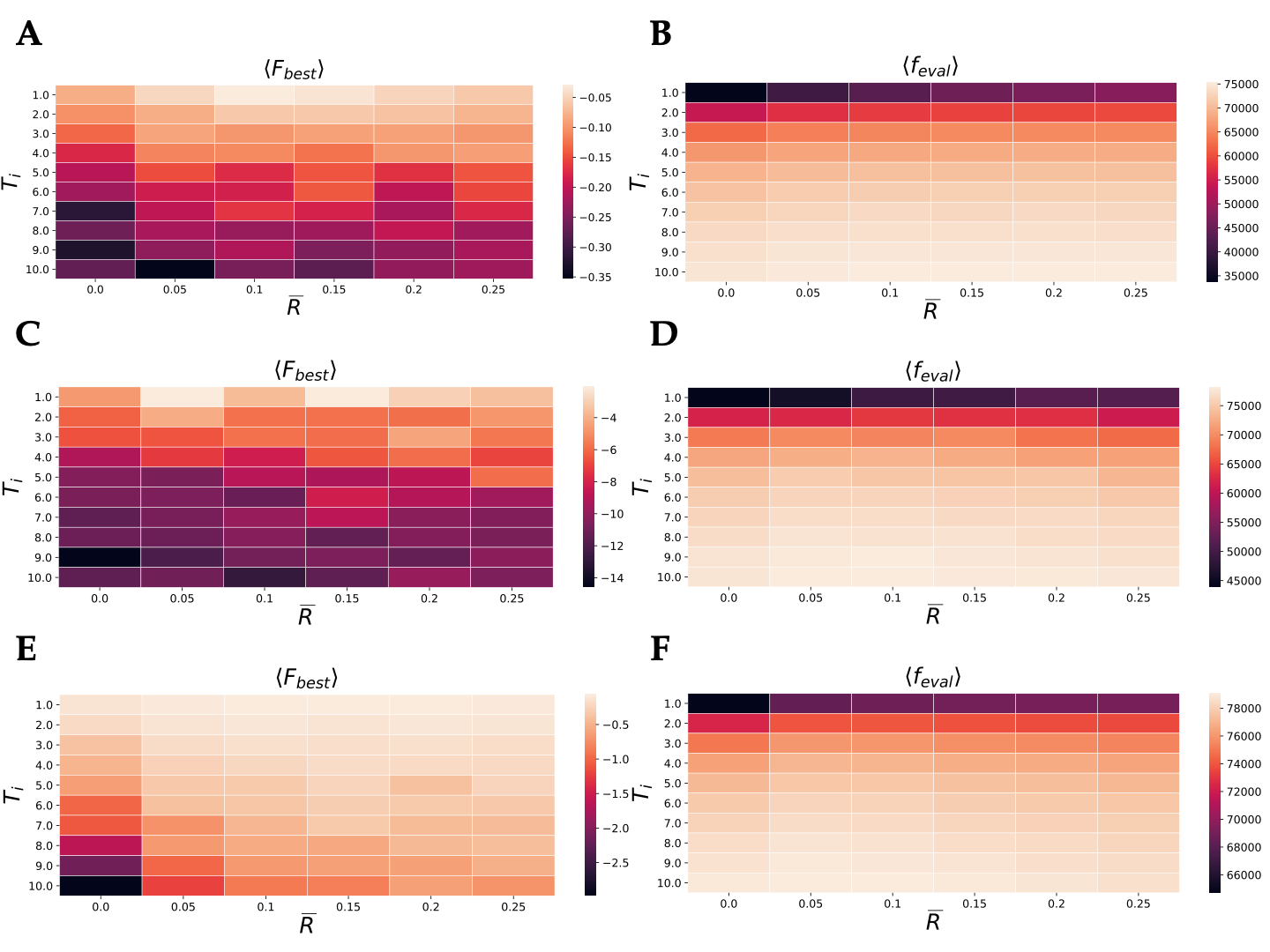}
\caption{
\textbf{Enhanced simulated annealing (ESA) hyperparameter scan: \textit{nnb} moveset.}
(A) A scan over ESA hyperparameters ($nnb$ moveset, linear cooling schedule) for the Rastrigin function:
the expected rate of fitness gain per step $\xbar{R}$ and the initial temperature $T_i$ (the final temperature is set to $T_f = 0.001$ in all plots).
Each cell in the heatmap represents best fitness values found in each run, averaged over $50$ independent runs with $l_\mathrm{tot} = 10^5$ steps each and randomly chosen starting states.
(B) Same as A but with the average taken over the number of fitness function evaluations (unique fitness function calls) in each ESA run.
(C) Same as A but for the Ackley function.
(D) Same as B but for the Ackley function.
(E) Same as A but for the Griewank function.
(F) Same as B but for the Griewank function.
}
\label{fig:SAruns}
\end{figure}

%%%%%%
\begin{figure}[!htb]
\centering
%\hspace*{-2.5cm}
\includegraphics[scale=0.60]{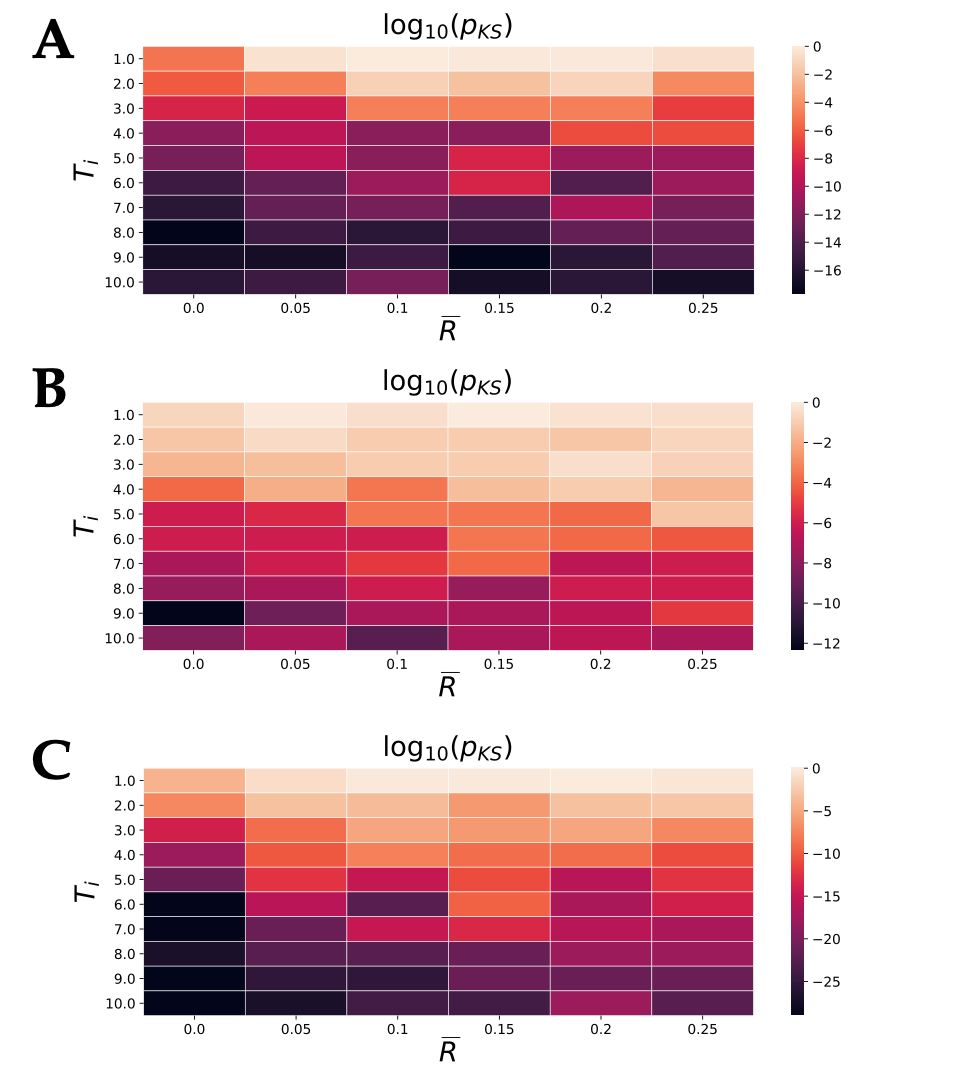}
\caption{
\textbf{Kolmogorov-Smirnov (KS) p-value analysis of the ESA hyperparameter settings: \textit{nnb} moveset.}
(A) One-sided KS tests were used to investigate the significance of the differences between two distributions of best-fitness values: the heatmap cell with the highest average of best-fitness values
(over $50$ independent ESA runs) vs. every other cell. Low p-values ($< 0.05$) indicate that the cell with the highest average has a significantly better distribution of best-fitness values than the other cell.
High p-values ($\ge 0.05$) indicate that the cell with the highest average has a distribution of best-fitness values that is either indistinguishable from, or worse than the distribution in the other cell.
ESA was run on the Rastrigin function with the $nnb$ moveset and a linear cooling schedule (Fig.~\ref{fig:SAruns}A).
(B) Same as A but for the Ackley function (Fig.~\ref{fig:SAruns}C).
(C) Same as A but for the Griewank function (Fig.~\ref{fig:SAruns}E).
}
\label{fig:SAruns:pKSself}
\end{figure}

%%%%%%
\begin{figure}[!htb]
\centering
%\hspace*{-2.5cm}
\includegraphics[scale=0.55]{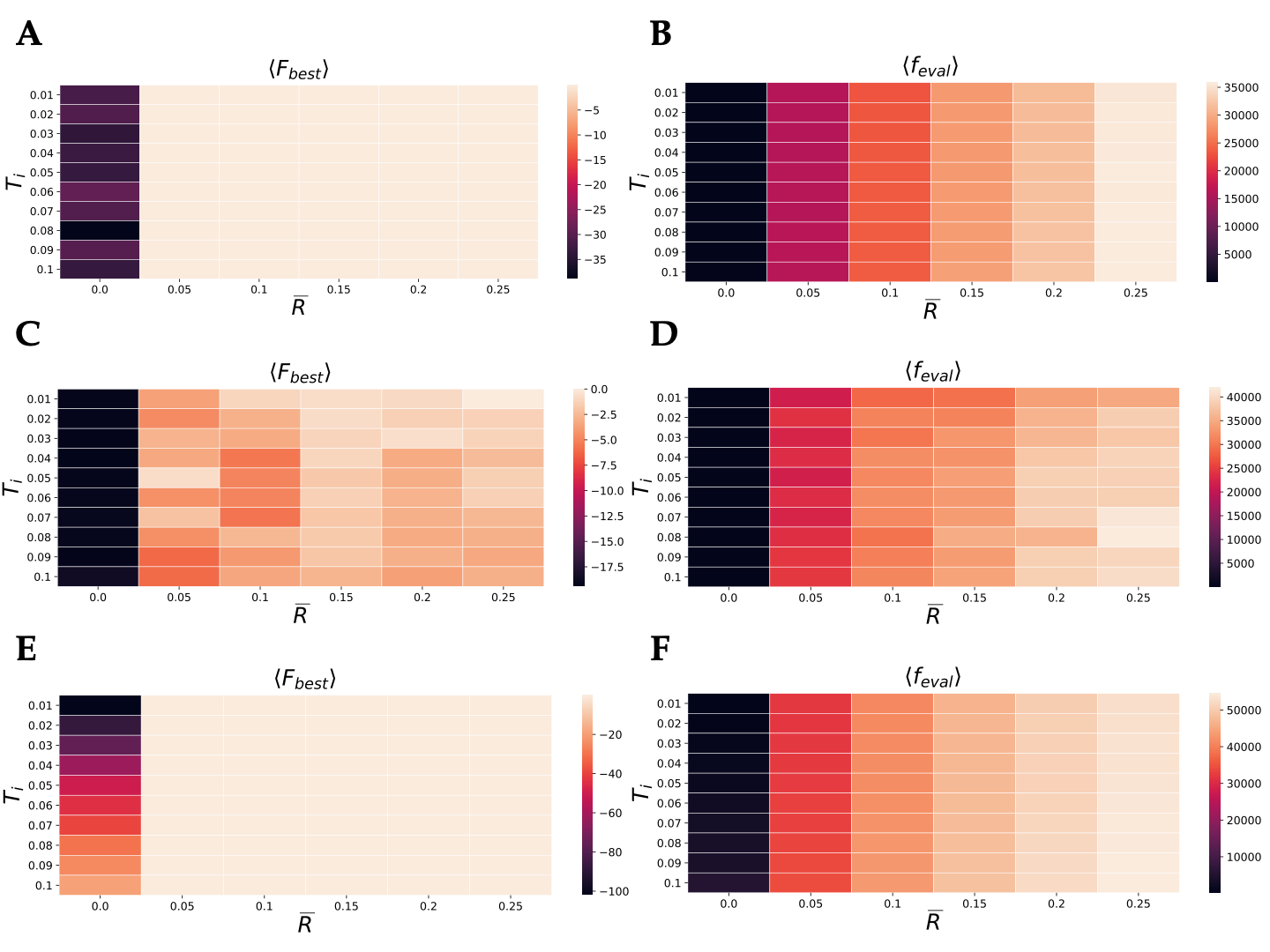}
\caption{
\textbf{Enhanced simulated annealing (ESA) hyperparameter scan: low $T_i$, \textit{nnb} moveset.}
Same as Fig.~\ref{fig:SAruns} but with suboptimally low initial temperatures $T_i$ employed with a linear cooling schedule.
}
\label{fig:SAruns:lowTi}
\end{figure}

%%%%%%
\begin{figure}[!htb]
\centering
%\hspace*{-2.5cm}
\includegraphics[scale=0.55]{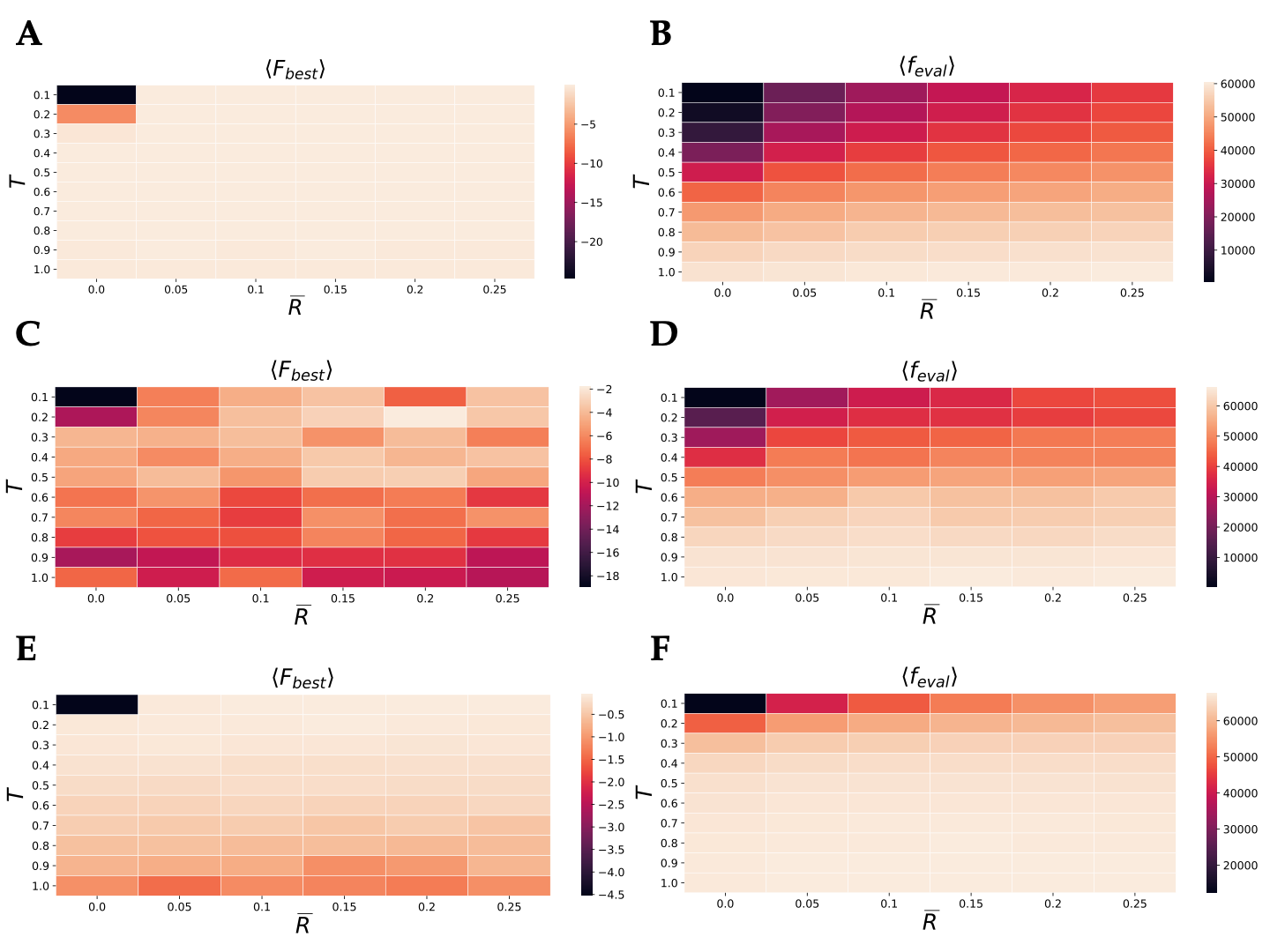}
\caption{
\textbf{Enhanced stochastic hill climbing (ESHC) hyperparameter scan: \textit{nnb} moveset.}
(A) A scan over ESHC hyperparameters ($nnb$ moveset) for the Rastrigin function:
the expected rate of fitness gain per step $\xbar{R}$ and the temperature $T$.
Each cell in the heatmap represents best fitness values found in each run, averaged over $50$ independent runs with $l_\mathrm{tot} = 10^5$ steps each and randomly chosen starting states.
(B) Same as A but with the average taken over the number of fitness function evaluations (unique fitness function calls) in each ESHC run.
(C) Same as A but for the Ackley function.
(D) Same as B but for the Ackley function.
(E) Same as A but for the Griewank function.
(F) Same as B but for the Griewank function.
}
\label{fig:SHCruns}
\end{figure}

%%%%%%
\begin{figure}[!htb]
\centering
%\hspace*{-2.5cm}
\includegraphics[scale=0.60]{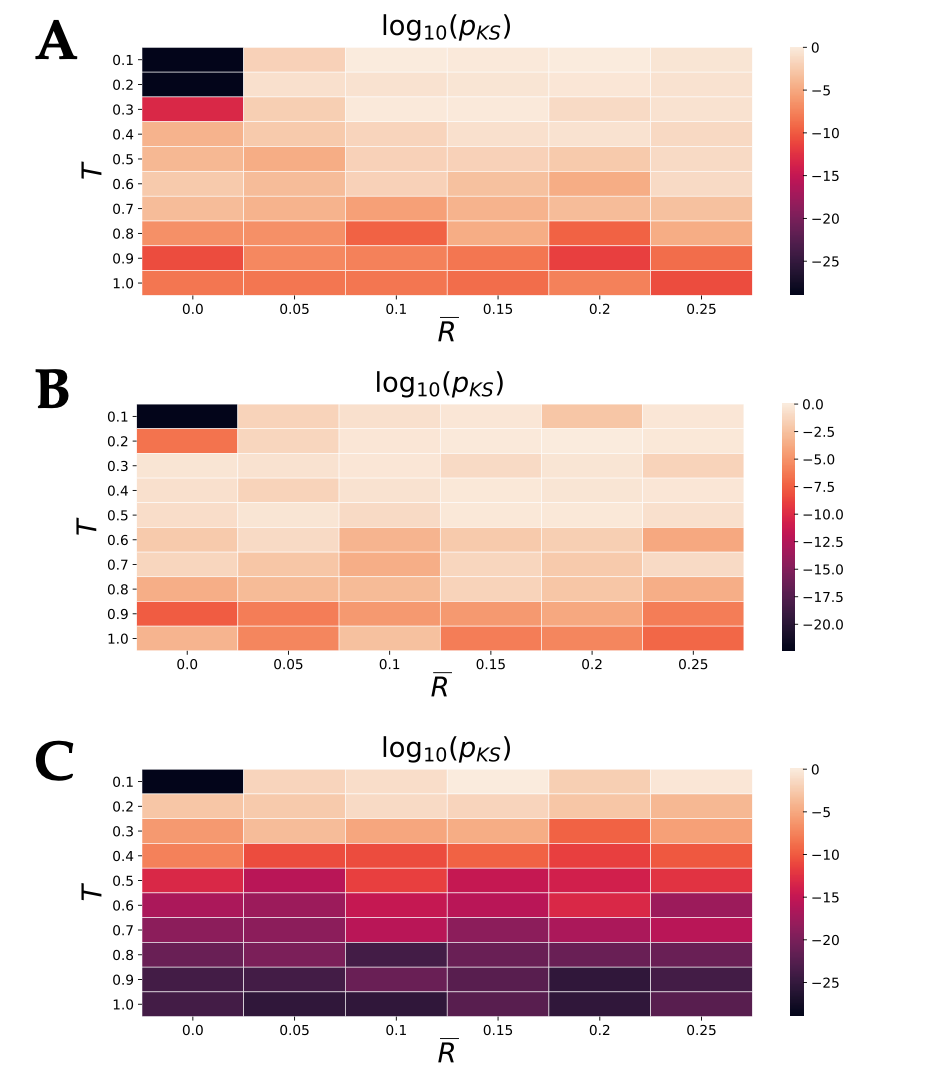}
\caption{
\textbf{Kolmogorov-Smirnov (KS) p-value analysis of the ESHC hyperparameter settings: \textit{nnb} moveset.}
(A) One-sided KS tests were used to investigate the significance of the differences between two distributions of best-fitness values: the heatmap cell with the highest average of best-fitness values
(over $50$ independent ESHC runs) vs. every other cell. Low p-values ($< 0.05$) indicate that the cell with the highest average has a significantly better distribution of best-fitness values than the other cell.
High p-values ($\ge 0.05$) indicate that the cell with the highest average has a distribution of best-fitness values that is either indistinguishable from, or worse than the distribution in the other cell.
ESHC was run on the Rastrigin function with the $nnb$ moveset (Fig.~\ref{fig:SHCruns}A).
(B) Same as A but for the Ackley function (Fig.~\ref{fig:SHCruns}C).
(C) Same as A but for the Griewank function (Fig.~\ref{fig:SHCruns}E).
}
\label{fig:SHCruns:pKSself}
\end{figure}

%%%%%%
\begin{figure}[!htb]
\centering
%\hspace*{-2.5cm}
\includegraphics[scale=0.55]{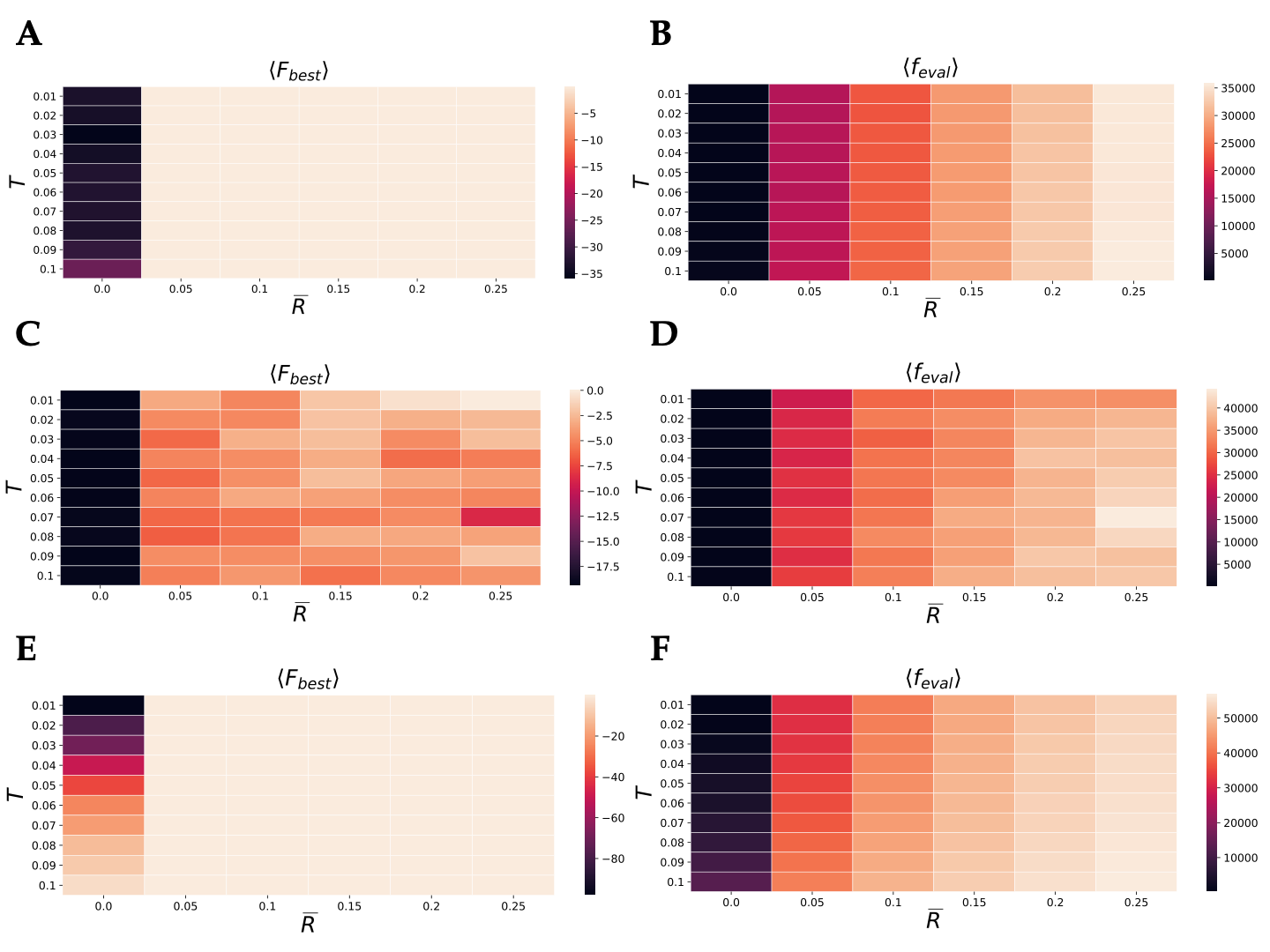}
\caption{
\textbf{Enhanced stochastic hill climbing (ESHC) hyperparameter scan: low $T$, \textit{nnb} moveset.}
Same as Fig.~\ref{fig:SHCruns} but with suboptimally low temperatures $T$.
}
\label{fig:SHCruns:lowT}
\end{figure}

%%%%%%
\begin{figure}[!htb]
\centering
%\hspace*{-2.5cm}
\includegraphics[scale=0.60]{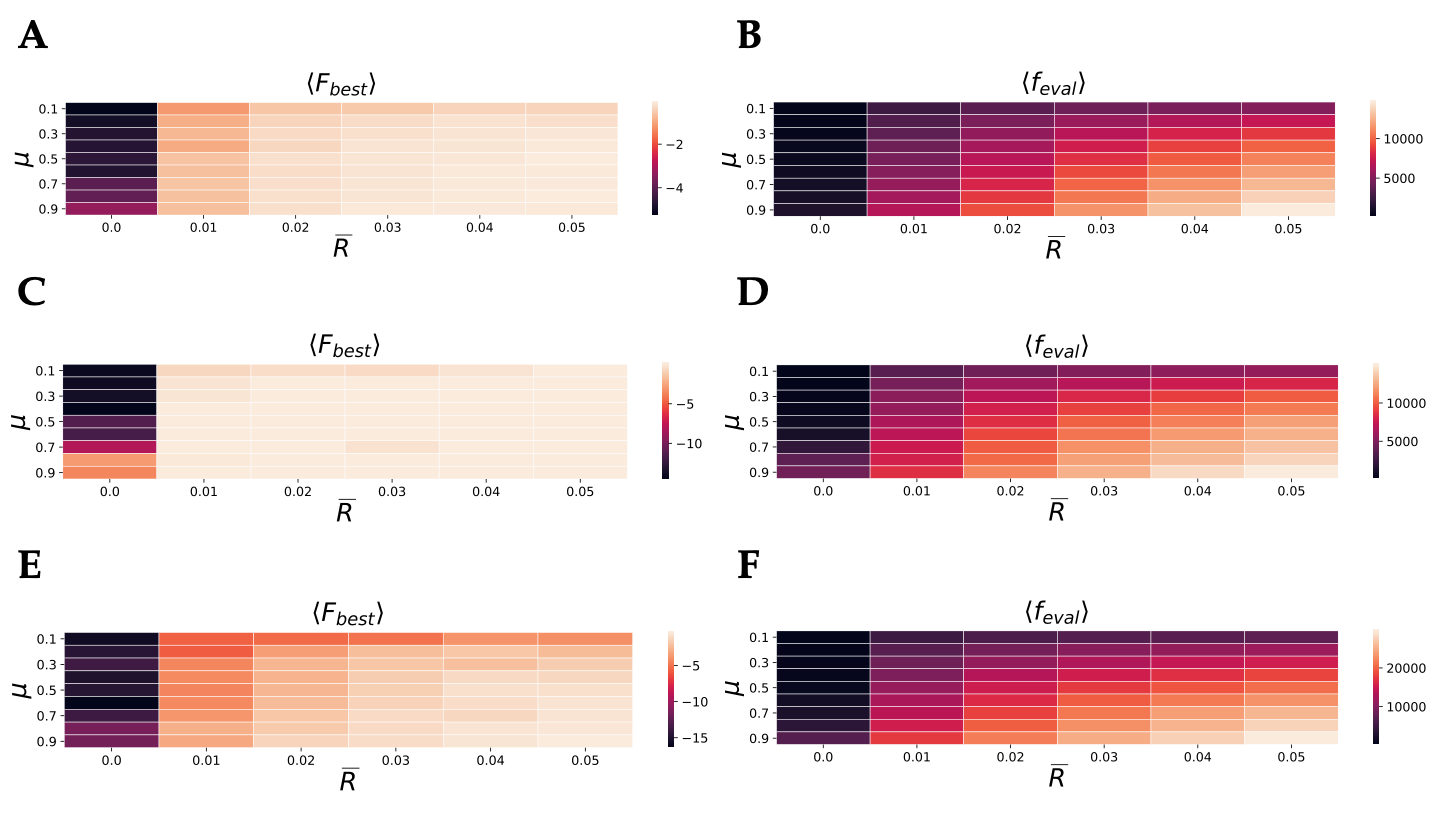}
\caption{
\textbf{Enhanced evolutionary algorithm (EEA) hyperparameter scan: \textit{nnb} moveset.}
(A) A scan over EEA hyperparameters ($nnb$ moveset) for the Rastrigin function:
the expected rate of fitness gain per step $\xbar{R}$ and the mutation rate $\mu$ (the crossover rate is set to $r_x = 0.2$ and the population size to $N_\text{pop} = 50$ in all plots).
Each cell in the heatmap represents best fitness values found in each run, averaged over $50$ independent runs with $l_\mathrm{tot} = 2000$ steps each and randomly chosen starting states
(each EEA step involves evaluating fitness functions for all $50$ population members).
(B) Same as A but with the average taken over the number of fitness function evaluations (unique fitness function calls) in each EEA run.
(C) Same as A but for the Ackley function.
(D) Same as B but for the Ackley function.
(E) Same as A but for the Griewank function.
(F) Same as B but for the Griewank function.
}
\label{fig:EAruns}
\end{figure}

%%%%%%
\begin{figure}[!htb]
\centering
%\hspace*{-2.5cm}
\includegraphics[scale=0.60]{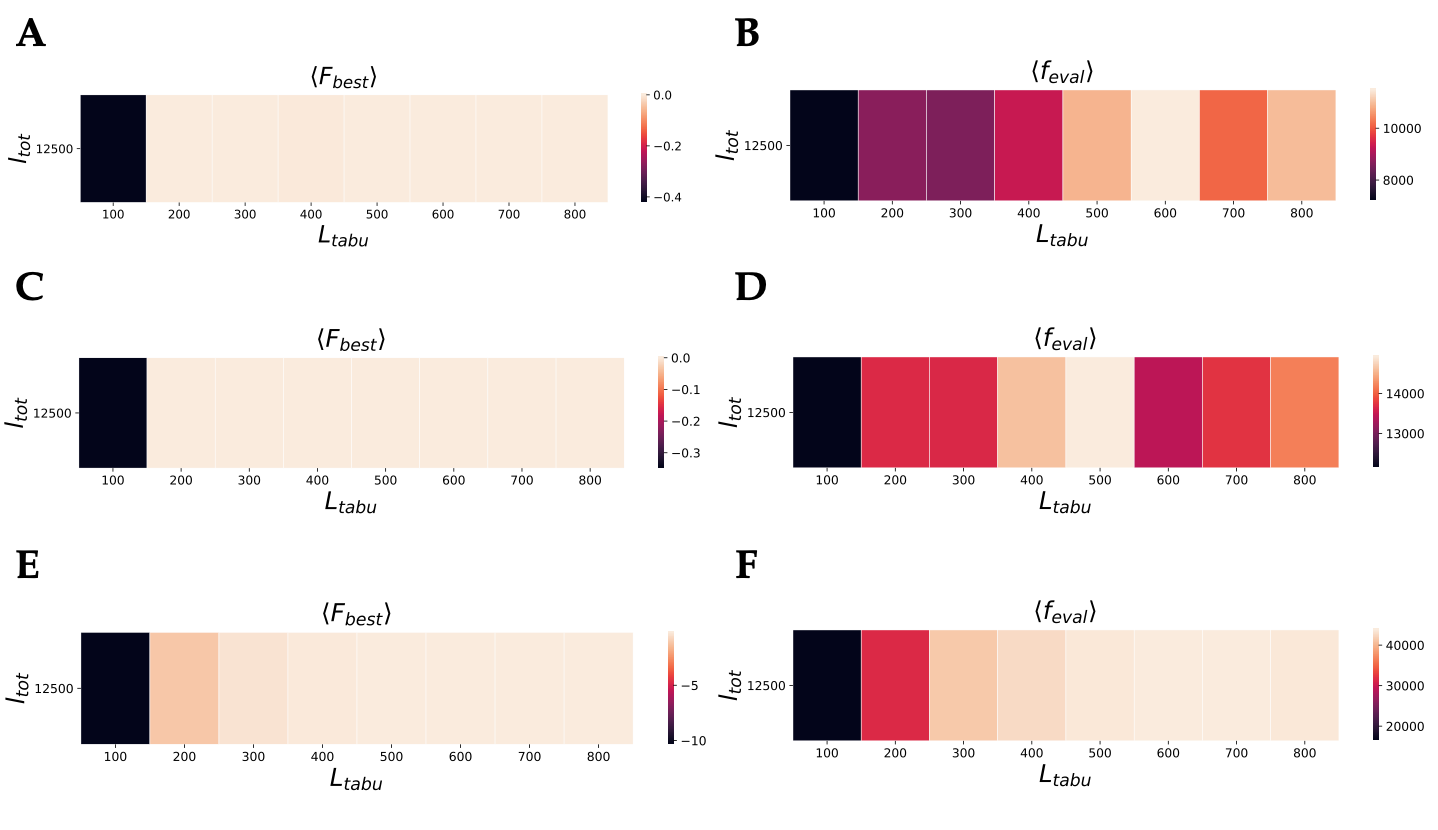}
\caption{
\textbf{Taboo search (TS) hyperparameter scan: \textit{nnb} moveset.}
(A) A scan over the length of the taboo list $L_\text{tabu}$ ($nnb$ moveset) for the Rastrigin function.
Each cell in the heatmap represents best fitness values found in each run, averaged over $50$ independent runs with $l_\mathrm{tot} = 12500$ steps each and randomly chosen starting states
(each TS step involves evaluating fitness functions for all $8$ nearest neighbors of the current node).
(B) Same as A but with the average taken over the number of fitness function evaluations (unique fitness function calls) in each TS run.
(C) Same as A but for the Ackley function.
(D) Same as B but for the Ackley function.
(E) Same as A but for the Griewank function.
(F) Same as B but for the Griewank function.
}
\label{fig:TSruns}
\end{figure}

%%%%%%
\begin{figure}[!htb]
\centering
%\hspace*{-2.5cm}
\includegraphics[scale=0.55]{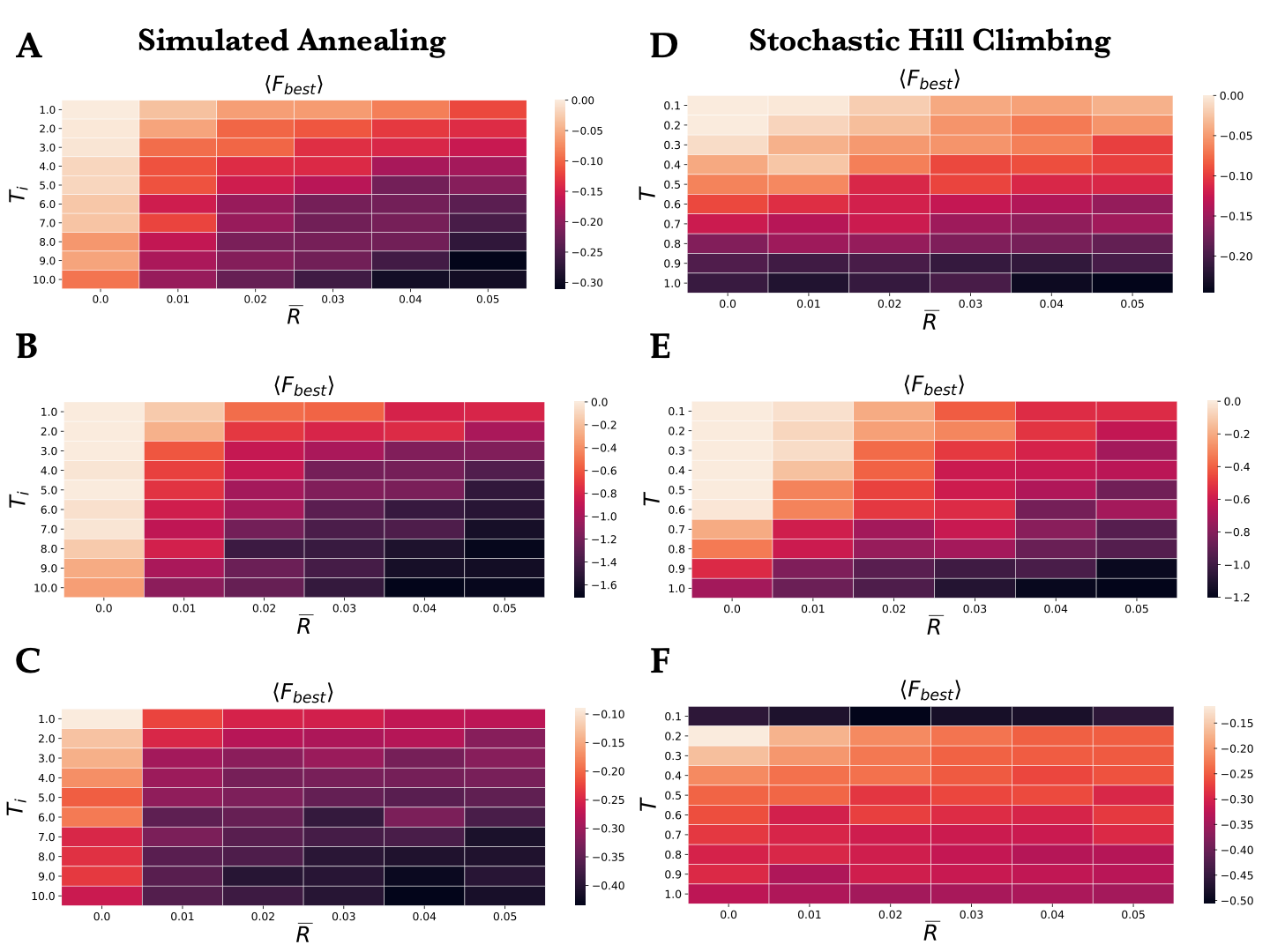}
\caption{
\textbf{ESA and ESHC hyperparameter scans: \textit{spmut} moveset.}
(A-C) Same as Fig.~\ref{fig:SAruns}A,C,E but for the \textit{spmut} moveset.
(D-F) Same as Fig.~\ref{fig:SHCruns}A,C,E but for the \textit{spmut} moveset.
}
\label{fig:SA_SHC_runs:spmut}
\end{figure}

%%%%%%
\begin{figure}[!htb]
\centering
%\hspace*{-2.5cm}
\includegraphics[scale=0.55]{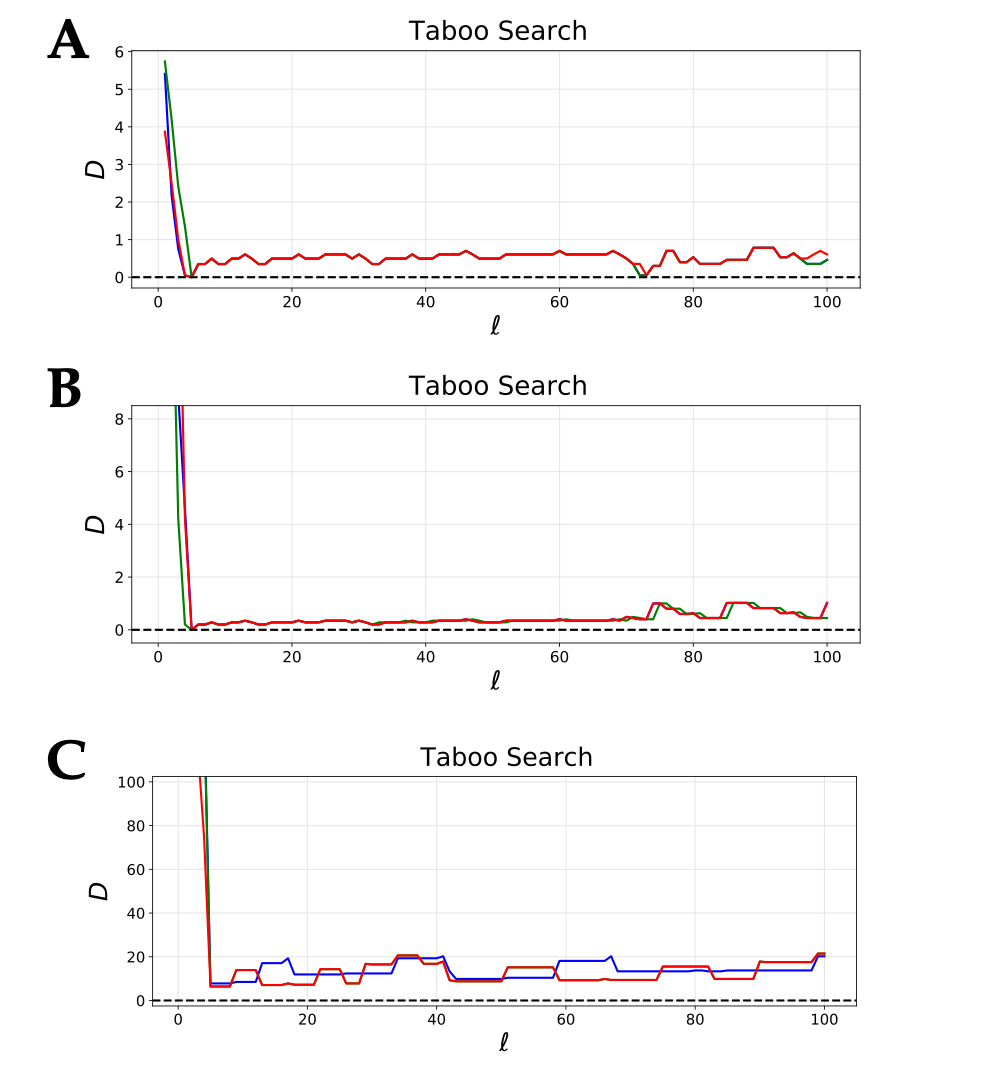}
\caption{
\textbf{Representative TS fitness trajectories: \textit{spmut} moveset.}
Shown is the $L_2$ distance $D$ between the current 4D state and the global maximum as a function of the number of steps $\ell$ for $3$ randomly chosen
TS trajectories.
TS was run with $L_\text{tabu} = 100$.
(A) Rastrigin function.
(B) Ackley function.
(C) Griewank function.
}
\label{fig:Dtraj:spmut}
\end{figure}

\begin{figure}[!htb]
\centering
%\hspace*{-2.5cm}
\includegraphics[scale=0.60]{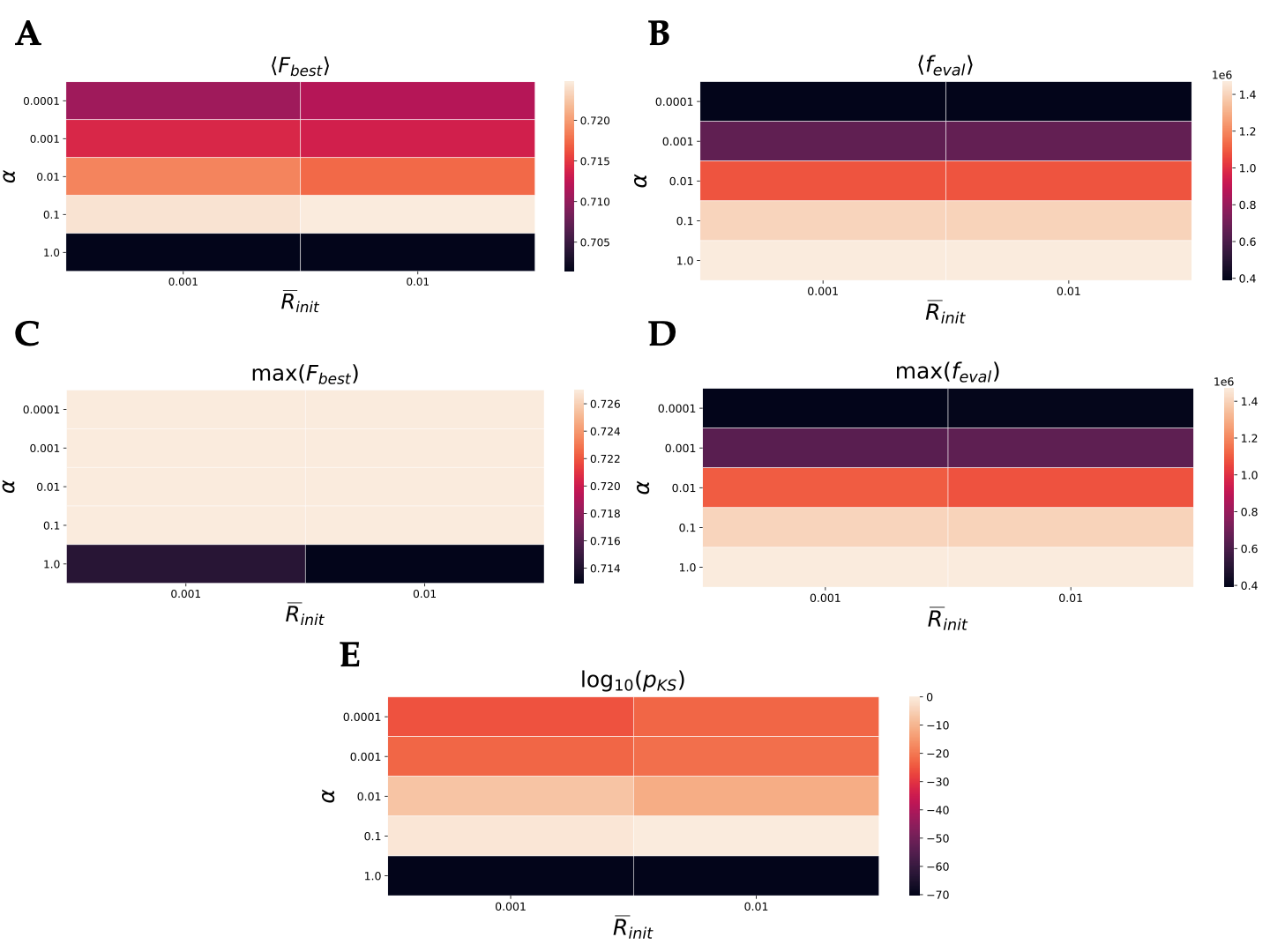}
\caption{
\textbf{SmartRunner (SR) hyperparameter scan: SK model with $\mathbf{N=200}$ spins.}
(A) A scan over SmartRunner hyperparameters ($l_\mathrm{max} = 2$): the initial value of the expected rate of fitness gain per step $\xbar{R}_\mathrm{~\!\!init}$ and the level of optimism $\alpha$.
Each cell in the heatmap represents the best fitness values found in each run, averaged over $100$ independent runs with $l_\mathrm{tot} = 1.5 \times 10^6$ steps each and randomly chosen starting states.
(B) Same as A but with the average taken over the number of fitness function evaluations (unique fitness function calls) in each run.
(C) Same as A but with the globally best fitness value obtained over all $100$ runs shown instead of the average.
(D) Same as B but with the number of fitness function evaluations corresponding to the globally best run from C shown instead of the average over $100$ runs.
(E) One-sided KS tests used to investigate the significance of the differences between two distributions of best-fitness values: the heatmap cell with the highest average of best-fitness values
in panel A vs. every other cell. Low p-values ($< 0.05$) indicate that the cell with the highest average has a significantly better distribution of best-fitness values than the other cell.
High p-values ($\ge 0.05$) indicate that the cell with the highest average has a distribution of best-fitness values that is either indistinguishable from, or worse than the distribution in the other cell.
}
\label{fig:SK200scan}
\end{figure}

%%%%%%
\begin{figure}[!htb]
\centering
%\hspace*{-2.5cm}
\includegraphics[scale=0.60]{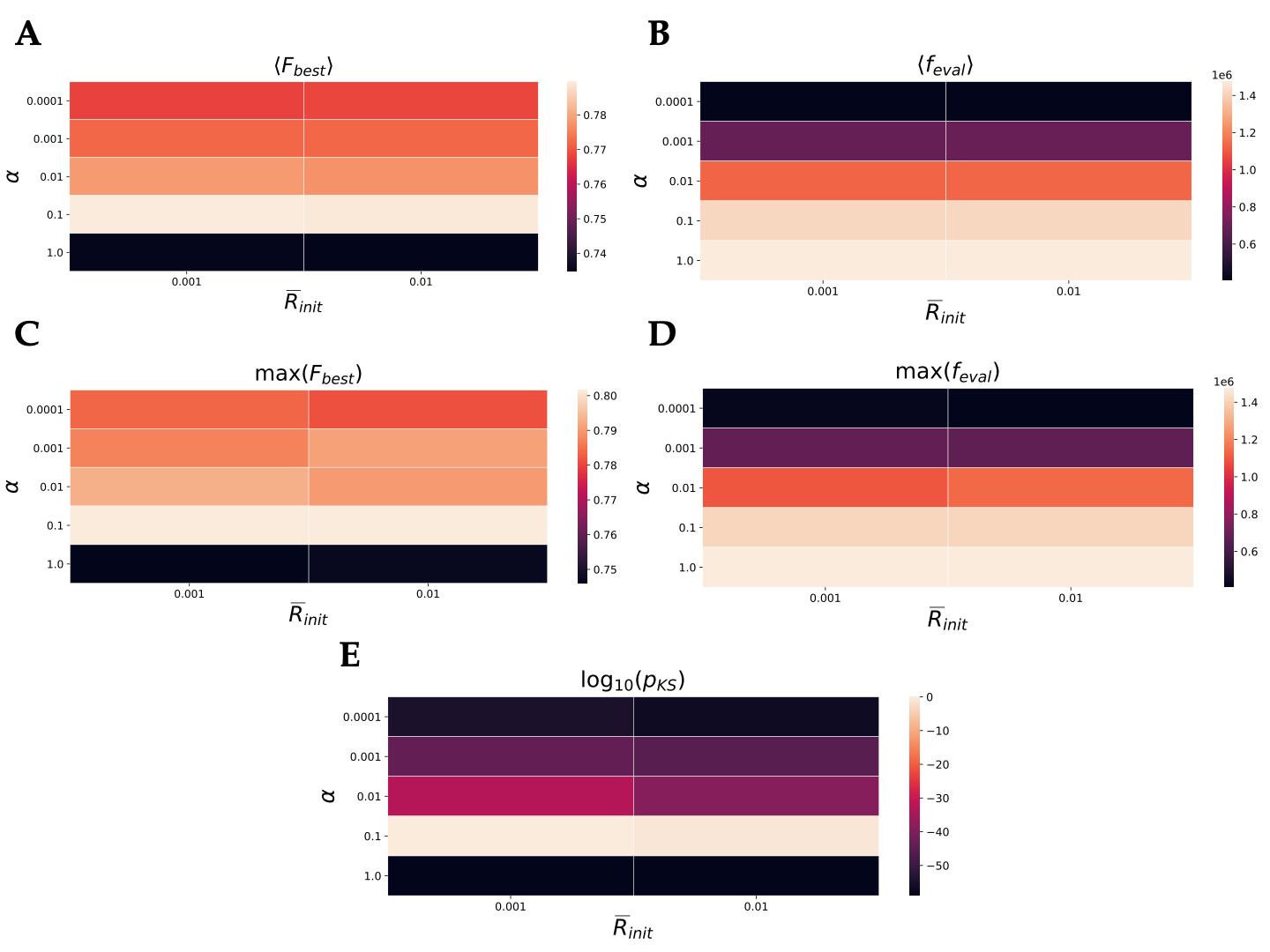}
\caption{
\textbf{SmartRunner (SR) hyperparameter scan: NK model with $\mathbf{N=200}$ sites and $\mathbf{K=8}$ nearest neighbors.}
Same as Fig.~\ref{fig:SK200scan} but for the NK model.
}
\label{fig:NK200.8scan}
\end{figure}

%%%%%%
\begin{figure}[!htb]
\centering
%\hspace*{-2.5cm}
\includegraphics[scale=0.70]{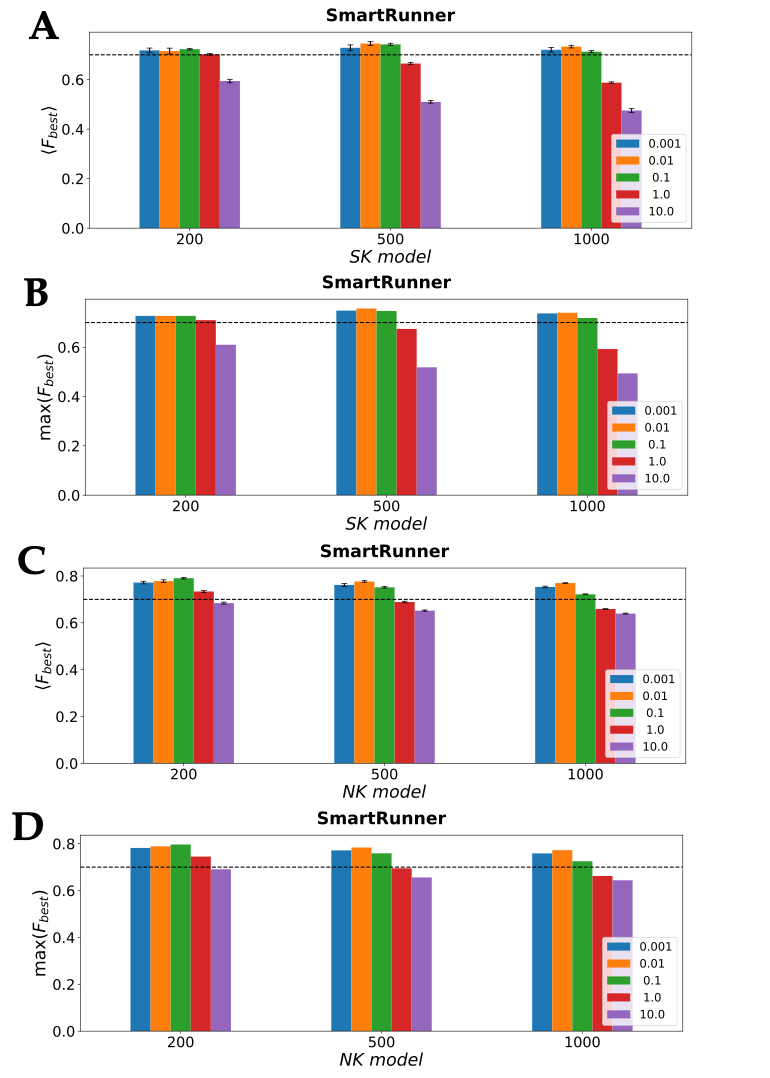}
\caption{
\textbf{SmartRunner (SR) hyperparameter scan: SK and NK models.}
The SK models have $N=200$, 500 and 1000 spins, while the NK models have $N=200$, 500 and 1000 sites, with $K=8$ randomly chosen intra-sequence couplings per site.
All SR runs were carried out with $l_\mathrm{max} = 2$ and $\xbar{R}^\mathrm{init} = 0.01$, and with
the total number of steps $l_\mathrm{tot} = 1.5 \times 10^6, 10^6, 5 \times 10^5$ for the models with 200, 500 and 1000 spins/sites, respectively.
(A) SK model: shown are the best fitness values found in each run, averaged over $10$ independent runs with randomly chosen starting states.
Error bars show standard deviations.
(B) SK model: shown are the globally best fitness values obtained over all $10$ runs.
(C) Same as A but for the NK model.
(D) Same as B but for the NK model.
The dashed horizontal line is drawn at $0.7$ to guide the eye. 
}
\label{fig:SKscan:3models}
\end{figure}

\clearpage

%\newpage
%\section*{Supplementary Tables}

%\section*{Acknowledgements}

\bibliographystyle{nar}
\bibliography{sample,optimization}